\documentclass[12pt,a4paper]{article}
\usepackage{amsmath} 
\usepackage{amssymb}
\usepackage{graphicx,epsfig}
\usepackage[mathscr]{eucal}
\usepackage{color} 
\usepackage[numbers,sort&compress]{natbib}
\usepackage[english]{babel}
\usepackage{booktabs}

\usepackage{tabularx}

\usepackage{setspace}
\usepackage{xspace}

\usepackage{multirow}

\newcommand{\mmumu}{m_{\mu\mu}\xspace}
\newcommand{\ymumu}{y_{\mu\mu}\xspace}
\newcommand{\yll}{y_{\mu\mu}\xspace}
\newcommand{\mll}{m_{\ell\ell}\xspace}
\newcommand{\afb}{A_{FB}(\mll)\xspace}
\newcommand{\seffl}{\sin^{2}\theta^{\ell}_{\rm{eff}}\xspace}

\newcommand\as{\alpha_{\mathrm{S}}}

\def\to{\rightarrow}

\usepackage{scalefnt,pstricks}
\definecolor{blue}{rgb}{0., 0., 1.0}

\def\rcut{r_{\rm cut}}

\def\muf{{\mu^{}_{F}}}

\def\mur{{\mu^{}_{R}}}

\newcommand\Matrix{{\sc Matrix}\xspace}
\newcommand\Munich{{\sc Munich}\xspace}
\newcommand\OpenLoops{{\sc OpenLoops}\xspace}
\newcommand\Recola{{\sc Recola}\xspace}

\usepackage{array}
\newcolumntype{L}[1]{>{\raggedright\let\newline\\\arraybackslash\hspace{0pt}}m{#1}}
\newcolumntype{C}[1]{>{\centering\let\newline\\\arraybackslash\hspace{0pt}}m{#1}}
\newcolumntype{R}[1]{>{\raggedleft\let\newline\\\arraybackslash\hspace{0pt}}m{#1}}

\usepackage[colorlinks=true,allcolors={blue!70!black}]{hyperref}

\begin{document} 
\begin{titlepage}
\begin{flushright}
ZU-TH 56/24\\
CERN-TH-2024-199\\
TIF-UNIMI-2024-18
\end{flushright}

\renewcommand{\thefootnote}{\fnsymbol{footnote}}

\begin{center}
  {\Large \bf Mixed QCD--EW corrections\\[0.3cm] to the neutral-current Drell--Yan process}
\end{center}

\par \vspace{1mm}
\begin{center}

  {\bf Tommaso Armadillo${}^{(a,b)}$, Roberto Bonciani${}^{(c,d)}$, Luca Buonocore${}^{(e)}$,\\[0.2cm]  Simone Devoto${}^{(f)}$, Massimiliano Grazzini${}^{(g)}$, Stefan Kallweit${}^{(g)}$, Narayan Rana${}^{(h)}$,}\\[0.2cm] and {\bf Alessandro Vicini${}^{(b)}$}\\[0.2cm] 
  \vspace{5mm}

${}^{(a)}$ Centre for Cosmology, Particle Physics and Phenomenology (CP3), Universite catholique de Louvain, Chemin du Cyclotron, 2, B-1348 Louvain-la-Neuve, Belgium\\[0.2cm]

${}^{(b)}$Dipartimento di Fisica “Aldo Pontremoli”, University of Milano and INFN, Sezione di Milano, I-20133 Milano, Italy\\[0.2cm]

${}^{(c)}$ 
Dipartimento di Fisica e Astronomia, Universit\`a di Firenze, I-50019 Sesto Fiorentino, Italy\\[0.2cm]
${}^{(d)}$Dipartimento di Fisica, Universit\`a di Roma “La Sapienza” and INFN, Sezione di Roma, I-00185 Roma, Italy\\[0.2cm]
${}^{(e)}$ CERN, Theoretical Physics Department, CH-1211 Geneva 23, Switzerland\\[0.2cm]
${}^{(f)}$ Department of Physics and Astronomy, Ghent University, 9000 Ghent, Belgium\\[0.2cm]
${}^{(g)}$ Physik Institut, Universit\"at Z\"urich, CH-8057 Z\"urich, Switzerland\\[0.2cm]
${}^{(h)}$School of Physical Sciences, National Institute of Science Education and Research,\\ An OCC of
Homi Bhabha National Institute, 752050 Jatni, India\\[0.2cm]

\end{center}

\par \vspace{2mm}
\begin{center} {\large \bf Abstract} 

\end{center}
\begin{quote}
\pretolerance 10000

We report on the complete computation of the mixed QCD--electroweak corrections to the neutral-current Drell--Yan process. Our calculation holds in the entire range of dilepton invariant masses.
We present phenomenological results for several kinematical distributions in the case of bare muons both in the resonant region and for high invariant masses. We also consider the forward--backward asymmetry, which is a key observable to measure the weak mixing angle.
 We finally extend our calculation to dressed leptons and compare our results in the massless limit to those available in the literature. 

\end{quote}

\vspace*{\fill}
\begin{flushleft}
December 2024
\end{flushleft}
\end{titlepage}

\vskip 0.5cm

\section{Introduction}

The Drell--Yan (DY) process~\cite{Drell:1970wh} has played a crucial role in the historical development of the theory of the strong and electroweak (EW) interactions.
It corresponds to the inclusive hadroproduction of a lepton pair through an off-shell vector boson.
The production rates are large and the experimental signatures are clean,
given the presence of at least one lepton with large transverse momentum in the final state, and the absence of colour flow between initial and final state.

The DY process is crucial for
the precision extraction of the SM parameters, such as proton parton distribution functions (PDFs) (see e.g.\ Ref.~\cite{Amoroso:2022eow} and references therein), 
the $W$ boson mass~\cite{ATLAS:2017rzl,LHCb:2021bjt,ATLAS:2024erm,CMS:2024nau}, the weak mixing angle~\cite{Aaltonen:2018dxj,LHCb:2015jyu,ATLAS:2018gqq,CMS:2024aps} and the strong coupling constant $\as$ (see e.g.\ Refs.~\cite{Ball:2018iqk,ATLAS:2023lhg}).

The DY process was one of the first hadronic reactions for which radiative corrections in the strong and EW couplings $\as$ and $\alpha$ were computed.
The classic calculations of the next-to-leading-order~(NLO)~\cite{Altarelli:1979ub}
and next-to-next-to-leading-order~(NNLO)~\cite{Hamberg:1990np,Harlander:2002wh} corrections to the total cross section in QCD were followed by
(fully) differential NNLO computations including the leptonic decay of the vector boson~\cite{Anastasiou:2003yy,Anastasiou:2003ds,Melnikov:2006kv,Catani:2009sm,Catani:2010en}.
The complete EW corrections for $W$ production have been computed in Refs.~\cite{Dittmaier:2001ay,Baur:2004ig,Zykunov:2006yb,Arbuzov:2005dd,CarloniCalame:2006zq},
and for $Z$ production in Refs.~\cite{Baur:2001ze,Zykunov:2005tc,CarloniCalame:2007cd,Arbuzov:2007db,Dittmaier:2009cr}.
Recently, the next-to-next-to-next-to-leading-order~(N$^3$LO) QCD radiative calculations for the inclusive production of a virtual photon~{\cite{Duhr:2020seh,Chen:2021vtu}} and of a $W$ boson~\cite{Duhr:2020sdp} have been completed, and also computations of fiducial cross sections at this order were performed~\cite{Camarda:2021ict,Chen:2022cgv,Neumann:2022lft,Campbell:2023lcy}.

The mixed QCD--QED corrections to the inclusive production of an on-shell $Z$ boson were obtained
in Ref.~\cite{deFlorian:2018wcj} through an abelianisation procedure from the NNLO QCD results~\cite{Hamberg:1990np,Harlander:2002wh}, and later
extended in Ref.~\cite{Cieri:2020ikq} to the fully differential level for off-shell $Z$ boson production and decay into a pair of neutrinos, thereby avoiding final-state radiation.
A similar calculation was carried out in Ref.~\cite{Delto:2019ewv} in an on-shell approximation for the $Z$ boson, but including the factorised NLO QCD corrections to $Z$ production and the NLO QED corrections to the leptonic $Z$ decay.
Complete ${\cal O}(\as\alpha)$ computations for the production of on-shell $Z$ and $W$ bosons have been presented
in Refs.~\cite{Bonciani:2016wya,Bonciani:2019nuy,Bonciani:2020tvf,Buccioni:2020cfi,Behring:2020cqi,Bonciani:2021iis}.
Beyond the on-shell approximation, important results have been obtained in the {\it pole} approximation (see e.g. Ref.~\cite{Denner:2019vbn} and references therein).
In this approximation the cross section around the $W$ or $Z$ resonance is systematically expanded so as to split the radiative corrections
into well-defined, gauge-invariant contributions.
Such method has been used in Refs.~\cite{Dittmaier:2014qza,Dittmaier:2015rxo} to evaluate the dominant part of the mixed QCD--EW corrections in the resonant region.
The calculation in this approximation was recently completed in Ref.~\cite{Dittmaier:2024row}.

A first step beyond the pole approximation has been carried out in Ref.~\cite{Dittmaier:2020vra}, where
results for the ${\cal O}(n_F\as \alpha)$ contributions to the DY cross section were presented.
In Ref.~\cite{Buonocore:2021rxx} some of us presented a computation of the mixed QCD--EW corrections to the charged-current process
\mbox{$pp\to \ell \nu_\ell+X$}, where all contributions are evaluated exactly except for the finite part of the two-loop amplitude, which is evaluated in the pole approximation.
Recently, the exact evaluation of the two-loop amplitude for the QCD--EW corrections to this process has been presented in Ref.~\cite{Armadillo:2024nwk}.

The complete computation of the mixed QCD--EW corrections to the neutral-current process \mbox{$pp\to \ell^+\ell^-+X$} has been reported in Ref.~\cite{Bonciani:2021zzf} considering massive leptons, and in Ref.~\cite{Buccioni:2022kgy} for massless leptons. The calculation of Ref.~\cite{Bonciani:2021zzf} is based on the exact two-loop amplitudes presented in Ref.~\cite{Armadillo:2022bgm}, evaluated through the reduction to the master integrals of Ref.~\cite{Bonciani:2016ypc} and a semi-analytical implementation of the differential-equations method for their calculation~\cite{Moriello:2019yhu,Hidding:2020ytt,Armadillo:2022ugh}. The computation of Ref.~\cite{Buccioni:2022kgy}  is based on the two-loop amplitudes presented in Ref.~\cite{Heller:2020owb}, expressed in terms of the master integrals evaluated in analytic polylogarithmic form~\cite{Heller:2019gkq,Hasan:2020vwn}.

The role of the mixed NNLO QCD--EW corrections is central for a correct interpretation of high-precision DY data~\cite{Balossini:2008cs,Balossini:2009sa, Grazzini:2019jkl}. 
According to the factorization theorem, we can write the hadron level cross section as a convolution of the proton PDFs with a partonic cross section, and both factors feature QCD and EW radiative effects. 
Initial-state QED and mixed QCD--QED collinear singularities can be factorized from the partonic cross section and reabsorbed in the definition of the physical proton PDF~\cite{deFlorian:2015ujt,deFlorian:2016gvk}. The coupled DGLAP evolution, with QCD and QED splitting kernels, yields a non-trivial interplay which goes beyond the simple single-emission probability, with mixed higher-order effects automatically included~\cite{Xie:2021equ,Cridge:2023ryv,Barontini:2024dyb}. The presence of a photon density in the proton opens the possibility of new partonic scattering channels, necessary to complement the cross sections initiated by quarks and gluons. The relative contribution of the various channels changes, as a function of the invariant mass of the final state, and only an NNLO QCD--EW calculation can consistently account for their correct balance~\cite{Mangano:2016jyj,Manohar:2016nzj,Manohar:2017eqh}.  The availability of an exact NNLO QCD--EW calculation throughout the whole invariant-mass range, including both the $Z$ resonance and the large invariant-mass tail, is important to predict the shape of the kinematical distributions, and, in turn, to establish in a statistically significant way any discrepancy from the data.
In a distinct and complementary perspective,
the increased precision of the partonic results may help putting tighter constraints on the proton parameterization at large partonic $x$~\cite{Ball:2022qtp}.
Moreover, the knowledge of the NNLO QCD--EW corrections potentially allows us
to improve over the approximations used in shower Monte Carlo programs~(see e.g.\ Refs.~\cite{Barze:2013fru,Kallweit:2015dum}, Sec.~IV.1 of Ref.~\cite{Andersen:2016qtm}, and the benchmarking study of Ref.~\cite{Alioli:2016fum}),
which include only partial subsets of factorisable mixed QCD--EW corrections, and to reduce the remaining theoretical uncertainties.

In this paper we follow up on the earlier results presented in Ref.~\cite{Bonciani:2021zzf} and present a detailed study on the impact of mixed QCD--EW corrections on the neutral-current DY process at LHC energies.
We start by considering bare muons and examine the mixed QCD--EW corrections in both the resonant region and in the region of high invariant masses.
We then analyse the impact of radiative corrections on the forward--backward asymmetry.
We finally consider the case of dressed leptons. 
Here we show that our calculation can be extended to the massless limit, and we present a comparison of our results with
those of the massless calculation of Ref.~\cite{Buccioni:2022kgy}.

The paper is organised as follows. In Sec.~\ref{sec:calc} we provide some details of our calculation.
In Sec.~\ref{sec:pheno} we present our phenomenological results, starting from the resonant region (Sec.~\ref{sec:reso}) and then moving to the region of high invariant masses (Sec.~\ref{sec:high}). We further present results for the forward--backward asymmetry (Sec.~\ref{sec:asym}), and finally consider the massless limit by presenting a comparison with the results of Ref.~\cite{Buccioni:2022kgy} (Sec.~\ref{sec:massless}). Our results are summarised in Sec.~\ref{sec:summa}.

\section{Calculational details}
\label{sec:calc}

We consider the inclusive production of a pair of massive muons in proton--proton collisions,
\begin{equation}
\label{eq:proc}
pp\to \mu^+\mu^-+X\, .
\end{equation}
The theoretical predictions for this process can be obtained as a convolution of the parton distribution functions for the incoming protons with the hard-scattering partonic cross sections.
When QCD and EW radiative corrections are considered, the initial partons include quarks, anti-quarks, gluons and photons.

The differential cross section for the process in Eq.~(\ref{eq:proc}) can be written as
\begin{equation}
  \label{eq:exp}
  d{\sigma}=\sum_{m,n=0}^\infty d{\sigma}^{(m,n)}\, ,
\end{equation}
where \mbox{$d{\sigma}^{(0,0)}\equiv d{\sigma}_{\rm LO}$} is the Born level contribution and $d{\sigma}^{(m,n)}$ the ${\cal O}(\as^m\alpha^n)$
correction.
The mixed QCD--EW corrections correspond to the term \mbox{$m=n=1$} in this expansion and include double-real, real--virtual and purely virtual contributions.
The corresponding tree-level and one-loop scattering amplitudes are computed with \OpenLoops~\cite{Cascioli:2011va, Buccioni:2017yxi, Buccioni:2019sur} and \Recola~\cite{Actis:2016mpe,Denner:2017wsf,Denner:2016kdg}, the results being in complete agreement\footnote{To be precise, the results obtained with \OpenLoops and \Recola coincide pointwise in phase space when adopting the $G_\mu$ or $\alpha(M_Z)$ EW renormalization schemes. They differ in the case of the $\alpha(0)$ scheme because of the different treatment of the light-fermion contributions to the photon vacuum polarization.}. 
The two-loop amplitude~\cite{Armadillo:2022bgm} is computed exploiting the reduction of the
scalar integrals appearing in the
corresponding Feynman diagrams to master integrals~\cite{Bonciani:2016ypc,Heller:2019gkq} using $\mathtt{KIRA}$~\cite{Maierhoefer:2017hyi}; their subsequent computation uses the semi-analytic approach to the differential-equations method~\cite{Moriello:2019yhu} as implemented in the Mathematica package $\mathtt{SeaSyde}$~\cite{Armadillo:2022ugh}. The masses of the vector bosons are renormalized in the 
complex-mass scheme~\cite{Denner:2005fg} and
consistently kept complex-valued at each step of the computation.
Even when all the amplitudes have been computed, the completion of the calculation remains a formidable task. Indeed, double-real, real--virtual and purely virtual contributions are separately infrared divergent,
and a method to handle and cancel infrared singularities has to be worked out.
In this calculation we use a formulation of the $q_T$ subtraction formalism~\cite{Catani:2007vq} derived from the NNLO QCD computation of heavy-quark pair production~\cite{Catani:2019iny,Catani:2019hip,Catani:2020kkl} through an appropriate abelianisation procedure~\cite{deFlorian:2018wcj,Buonocore:2019puv}.
According to the $q_T$ subtraction formalism~\cite{Catani:2007vq}, $d{\sigma}^{(m,n)}$ can be evaluated as
\begin{equation}
\label{eq:master}
d{\sigma}^{(m,n)}={\cal H}^{(m,n)}\otimes d{\sigma}_{\rm LO}+\left[d\sigma_{\rm R}^{(m,n)}-d\sigma_{\rm CT}^{(m,n)}\right]\, .
\end{equation}
The first term in Eq.~(\ref{eq:master}) is obtained through a convolution (denoted by the symbol $\otimes$)
of the perturbatively computable function ${\cal H}^{(m,n)}$ and the LO cross section $d{\sigma}_{\rm LO}$,
with respect to the longitudinal-momentum fractions of the colliding partons.
The second term is the {\it real} contribution $d\sigma_{\rm R}^{(m,n)}$, where the charged leptons
are accompanied by additional QCD and/or QED radiation that produces a recoil with finite transverse momentum $q_T$.
For \mbox{$m+n=2$} such contribution can be evaluated by using the dipole subtraction formalism~\cite{Catani:1996jh,Catani:1996vz,Catani:2002hc,Kallweit:2017khh,Dittmaier:1999mb,Dittmaier:2008md,Gehrmann:2010ry,Schonherr:2017qcj}.
In the limit \mbox{$q_T\to 0$} the real contribution
$d\sigma_{\rm R}^{(m,n)}$ is divergent, since the recoiling radiation becomes soft and/or collinear to the initial-state partons.
Such divergence is cancelled by the counterterm $d\sigma_{\rm CT}^{(m,n)}$, which eventually makes the cross section in Eq.~(\ref{eq:master}) finite.\footnote{More precisely, the square bracket in Eq.~(\ref{eq:master}) is evaluated with a cut \mbox{$q_T/\mmumu>\rcut$} ($\mmumu$ being the invariant mass of the dimuon system), and then an extrapolation for \mbox{$\rcut\to 0$} is carried out (see Ref.~\cite{Grazzini:2017mhc} for more details).}

The required phase space generation and integration is carried out within the \Matrix framework~\cite{Grazzini:2017mhc}.
The core of \Matrix is the Monte Carlo program \Munich\footnote{\Munich, which is the abbreviation of “MUlti-chaNnel Integrator at Swiss~(CH) precision”, is an automated parton-level NLO\
generator by S. Kallweit.}, which contains a fully automated implementation of the dipole subtraction method for massless and massive partons
at NLO QCD~\cite{Catani:1996jh,Catani:1996vz,Catani:2002hc} and NLO EW~\cite{Kallweit:2017khh,Dittmaier:1999mb,Dittmaier:2008md,Gehrmann:2010ry,Schonherr:2017qcj}.
The $q_T$ subtraction method has been applied to several NNLO QCD computations for the production of colourless final-state systems (see Ref.~\cite{Grazzini:2017mhc} and references therein), and to heavy-quark pair production~\cite{Catani:2019iny,Catani:2019hip,Catani:2020kkl} and related processes~\cite{Catani:2022mfv,Buonocore:2022pqq,Buonocore:2023ljm,Devoto:2024nhl}, which correspond to the case \mbox{$m=2$}, \mbox{$n=0$}.
The method has also been used in Ref.~\cite{Buonocore:2019puv} to study NLO EW corrections for the DY process, which represents the case \mbox{$m=0$}, \mbox{$n=1$}.
In the last few years some of us have applied this method to the computation of mixed QCD--EW corrections for the charged-current DY process~\cite{Buonocore:2021rxx}, while first results
on the neutral-current DY process were presented in Ref.~\cite{Bonciani:2021zzf}.
The structure of the coefficients ${\cal H}^{(1,1)}$ and $d\sigma_{\rm CT}^{(1,1)}$ can be derived from those controlling
the NNLO QCD computation of heavy-quark pair production.
The initial-state soft/collinear and purely collinear contributions were already presented in Ref.~\cite{Cieri:2020ikq}.
The fact that the final state is colour neutral implies that final-state radiation is of pure QED origin.
Therefore, the purely soft contributions
have a simpler structure than the corresponding contributions entering the
NNLO QCD computation of heavy-quark pair production~\cite{Catani:2023tby}.

\section{Phenomenological results}
\label{sec:pheno}

Unless stated otherwise,
we work in the $G_\mu$ scheme with \mbox{$G_F=1.1663787\times 10^{-5}$\,GeV$^{-2}$} and set the \textit{on-shell} values of masses and widths 
to \mbox{$m_{W,{\rm OS}}=80.385$\,GeV}, \mbox{$m_{Z, {\rm OS}}=91.1876$\,GeV}, \mbox{$\Gamma_{W, {\rm OS}}=2.085$\,GeV}, \mbox{$\Gamma_{Z, {\rm OS}}=2.4952$\,GeV}.
Those values are translated to the corresponding \textit{pole} values \mbox{$m_{V}=m_{V,{\rm OS}}/\sqrt{1+\Gamma^2_{V,{\rm OS}}/m^2_{V{,\rm OS}}}$} and \mbox{$\Gamma_{V}=\Gamma_{V,{\rm OS}}/\sqrt{1+\Gamma^2_{V,{\rm OS}}/m_{V,{\rm OS}}^2}$}, \mbox{$V=W,Z$}, from which \mbox{$\alpha=\sqrt{2}\,G_F m_{W}^2(1-m_{W}^2/m_{Z}^2)/\pi$} is derived,
and we use the complex-mass scheme~\cite{Denner:2005fg} throughout.
The muon mass is fixed to \mbox{$m_\mu=105.658369$\,MeV}, and the pole masses of the top quark and the Higgs boson to \mbox{$m_t=173.07$\,GeV} and \mbox{$m_H=125.9$\,GeV}, respectively.
The CKM matrix is taken to be diagonal.
We work with \mbox{$n_f=5$} massless quark flavours and retain the exact top-mass dependence in all virtual and real--virtual
amplitudes associated with bottom-induced processes, except for the two-loop virtual
corrections,  where we neglect top-mass effects.
Given the smallness of the bottom-quark density, we estimate the corresponding error to be at the percent level of the computed correction.
We use the \texttt{NNPDF31$\_$nnlo$\_$as$\_$0118$\_$luxqed} set of parton distributions~\cite{Bertone:2017bme}, which is based on the LUXqed methodology~\cite{Manohar:2016nzj}
for the determination of the photon density.
If not stated otherwise, the central values of the renormalisation and factorisation scales are fixed to \mbox{$\mu_R=\mu_F=m_Z$} (and the corresponding value of $\as$ is set), while
theoretical uncentainties are estimated by the customary 7-point
scale variation, i.e.\ by varying $\mur$ and $\muf$ by a factor of two
around their central values with the constraint \mbox{$1/2 <\mur/\muf < 2$}.
We consider bare muons in the final state, unless stated otherwise.

Our results for the mixed QCD--EW corrections will be compared with those from an approach in which QCD
and EW corrections are assumed to completely factorise. Such approximation can be defined
as follows: for each bin of a distribution \mbox{$d\sigma/dX$}, the QCD correction, \mbox{$d\sigma^{(1,0)}/dX$}, and the EW correction restricted to
the $q{\bar q}$ channel, \mbox{$d\sigma^{(0,1)}_{q{\bar q}}/dX$}, are computed, and the factorised ${\cal O}(\as\alpha)$
correction is calculated as
\begin{equation}
\label{eq:fact}
\frac{d\sigma_{\rm fact}^{(1,1)}}{dX}=   
\left(\frac{d\sigma^{(1,0)}}{dX}\right)
\times
\left(\frac{d\sigma^{(0,1)}_{q{\bar q}}}{dX}\right)
\times
\left(\frac{d\sigma_{\rm LO}}{dX}\right)^{-1}\, .
\end{equation}
This approximation is justified if the dominant sources of QCD and EW corrections factorise with respect to the hard gauge-boson production subprocess (see the discussion in Ref.~\cite{Buonocore:2021rxx}).

\subsection{Resonant region}
\label{sec:reso}

We first study the impact of the mixed corrections on the bulk of the DY
cross section around the $Z$ peak. We consider proton--proton collisions at \mbox{$\sqrt{s}=13.6\,$TeV} and a
fiducial volume defined by staggered cuts\footnote{This combination of cuts is
  known to restore the quadratic dependence of the acceptance at small transverse
  momentum of the dilepton pair, see e.g.\ Refs~\cite{Grazzini:2017mhc,Salam:2021tbm}.} on the
transverse momenta of the positively/negatively charged leptons,
\begin{equation}\label{eq:ResonantFidCuts-staggered}
p_{T,\mu^{+}} > 27\,{\rm GeV}\,,\quad p_{T,\mu^{-}} > 25\,{\rm GeV}\,,\quad |y_{\mu}|<2.5,
\end{equation}
while the invariant mass of the dimuon pair $\mmumu$ fulfils
\begin{equation}
66~{\rm GeV}<\mmumu<116~{\rm GeV}\, . 
\end{equation}

In Table~\ref{tab:res}
we show our predictions for the
fiducial cross section with an increasing radiative content.
We start our discussion from the pure QCD predictions, labelled LO, NLO$_{\rm QCD}$ and NNLO$_{\rm QCD}$. The NLO QCD corrections
increase the fiducial cross section by about $7\%$ with respect to
the LO, their perturbative uncertainty, estimated through scale variations, being
about $4\%$. The NNLO QCD corrections are negative, and amount to about $-2\%$,
while the associated uncertainties are reduced to below the percent level.
N$^3$LO corrections with this setup are not available, but their impact is likely to exceed the NNLO scale uncertainties~\cite{Chen:2022cgv}.

Predictions including NLO EW effects are shown in the fourth and fifth rows of Table~\ref{tab:res}, and are labelled NLO$_{\rm EW}$ and
\mbox{NNLO$_{\rm QCD}$+NLO$_{\rm EW}$}, respectively, while our best prediction, which includes mixed QCD--EW corrections as well, is labelled \mbox{NNLO$_{\rm QCD+MIX}$+NLO$_{\rm EW}$}.
We see that NLO EW corrections are negative, about $-4\%$, and have a larger impact than what might be expected by
a naive coupling counting. Indeed, they are similar in size to the NLO QCD effects and, being negative, largely compensate the latter.
We observe that the scale dependence of the NLO$_{\rm EW}$ prediction is identical to that at LO, as might have been expected, given that the relevant amplitudes do not involve the QCD coupling $\as$.
The newly computed mixed QCD--EW corrections are small for this particular
setup, being below half a percent. They slightly reduce the fiducial cross
sections.
The small size of the mixed QCD--EW corrections is consistent with the fact that they do not reduce the scale dependence. 

\begin{table}[h!]
\small
\renewcommand{\arraystretch}{1.4}
  \centering
    \begin{tabular}{l | c c c }
     & $\sigma$ [pb]  &  $\sigma^{(i,j)}$ [pb] &  $\sigma^{(i,j)}/\sigma_{\rm LO}$
     \\ \midrule
LO & $735.73(3)^{+12.7\%}_{ -13.6\%}$ & $-$ & $-$
       \\ [1.5ex]
NLO$_{\rm QCD}$ & $790.23(4)^{+2.7\%}_{ -4.4\%}$ & $\sigma^{(1,0)}= +54.50(5)$ & $+7.4$\%\\[1.5ex]
NNLO$_{\rm QCD}$ & $774.4(7)^{+0.7\%}_{-0.7\%}$ & $\sigma^{(2,0)}=-15.8(5)\phantom{0}$ & $-2.2$\%\\ [1.5ex]

NLO$_{\rm EW}$ & $704.07(3)\phantom{.}^{+12.7\%}_{ -13.6\%}$ & $\sigma^{(0,1)}=-31.66(3)$ & $-4.3$\%\\ [1.5ex]

NNLO$_{\rm QCD}$+NLO$_{\rm EW}$ &
$742.7(5)^{+0.3\%}_{-0.7\%}$ & $-$ & $-$\\ [1.5ex]

NNLO$_{\rm QCD+MIX}$+NLO$_{\rm EW}$ & $739.6(5)^{\phantom{0}+0.6\%}_{\phantom{0}-0.6\%}$ & $\sigma^{(1,1)}=-3.1(2)$ & $-0.4$\%\\
\bottomrule
  \end{tabular} 
\caption{\label{tab:res} Fiducial cross section at the different perturbative orders and the corresponding corrections $\sigma^{(i,j)}$ as defined in Eq.~\eqref{eq:exp}.}

\end{table}

\begin{figure}
  \centering
  \includegraphics[height=0.43\textheight]{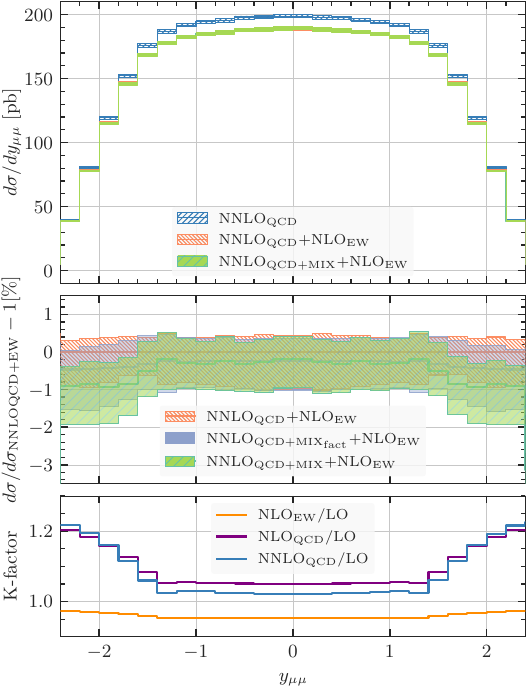}
  \hfill 
  \includegraphics[height=0.43\textheight]{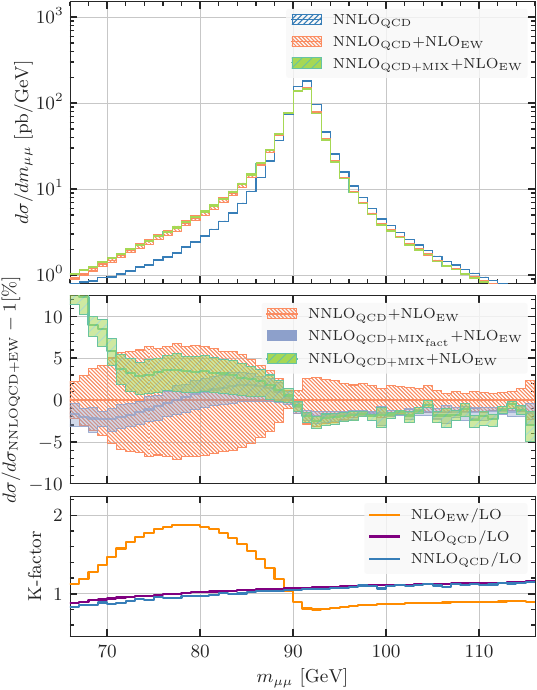}
  \caption{\label{fig:ResonantFid1} Predictions for the rapidity distribution (left) and the invariant-mass distribution (right) of the final-state muon pair, in the setup defined in Section \ref{sec:reso}. 
  NLO EW and mixed QCD--EW corrections are additively included on top of the NNLO QCD prediction. 
  The relative effects of the mixed corrections on top of the additive combination \mbox{NNLO QCD + NLO EW} are shown in the middle panel, comparing the exact mixed corrections to the approximate factorised ansatz defined in the text. 
  NLO EW as well as NLO and NNLO QCD $K$-factors are displayed in the bottom panel.}
\end{figure}

We now turn to differential distributions.
In Fig.~\ref{fig:ResonantFid1} (left) we show our results for the rapidity distribution $\ymumu$ of the dimuon pair.  The strongest shape effect can be observed at NLO QCD and to a lesser extent at NNLO QCD. The inclusion of the mixed QCD--EW corrections further distorts the shape of the distribution up to 1\% in the region \mbox{$|\ymumu|>1.5$}, and could be potentially relevant for a determination of the proton PDFs with 1\% precision
\cite{NNPDF:2024dpb}.

In Fig.~\ref{fig:ResonantFid1} (right), we consider the
invariant-mass distribution of the dimuon system. QCD radiative
corrections are generally mild for this observable, while the emission
of QED radiation off the final-state lepton pair has a significant
impact if the invariant mass is reconstructed from bare leptons, as we
do here. 
The inclusion of the NLO EW corrections indeed leads to a
shift of events, giving rise to the formation of the characteristic radiative
tail in the region on the left of the peak.\footnote{The effect is also present but reduced by approximately a factor of two when leptons and photons are recombined~\cite{CarloniCalame:2007cd}.}
Purely weak NLO corrections modify the strength of the $Z$-boson couplings to fermions and in turn, at LO, the squared matrix element of the $Z$-exchange diagram and the $\gamma$--$Z$ interference. They affect the region around the $Z$ resonance, with positive corrections reaching up to a few percent below the resonance and moderate negative corrections above the resonance~\cite{Dittmaier:2024row}.

Focusing on the impact of
the mixed QCD--EW corrections in the middle plot, we observe sizeable
effects and a non-trivial shape distortion. The correction is vanishing around
the peak, where it reaches the maximum slope, while it is rather flat,
positive and around $4\%$ in the region below the peak and rather
flat, negative and around $-2\%$ in the region above. We observe that
the prediction obtained with the factorised ansatz (see Eq.~(\ref{eq:fact})) works well in the
latter region, while it fails to describe the exact result below the $Z$ resonance, 
because of a non-trivial kinematical interplay between the production of the gauge boson and its subsequent decay into leptons, 
as it has been first pointed out in Ref.~\cite{Dittmaier:2015rxo}.
Furthermore, we notice that the
inclusion of the mixed corrections leads to a significant reduction of the
uncertainty band.

\begin{figure}
  \centering
  \includegraphics[height=0.43\textheight]{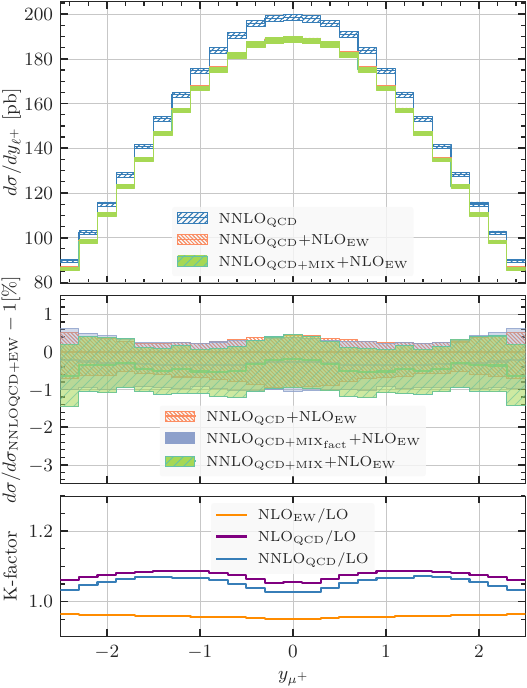}
  \hfill 
  \includegraphics[height=0.43\textheight]{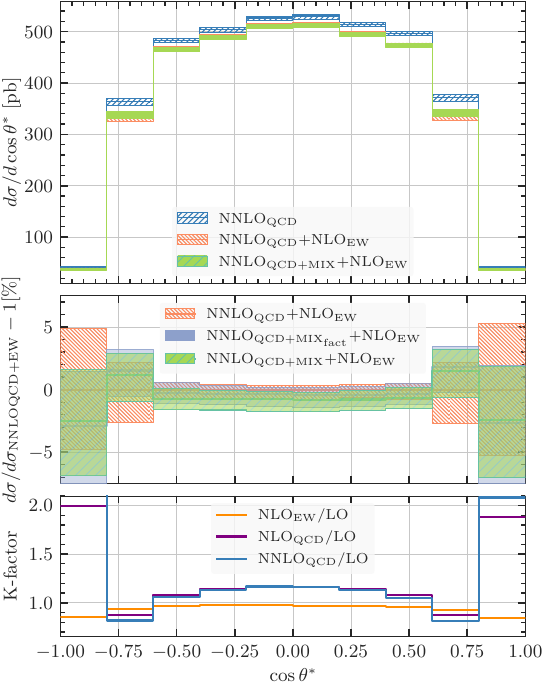}
  \caption{\label{fig:ResonantFid2} Same as Fig.~\ref{fig:ResonantFid1} for the rapidity
    distribution of the anti-muon (left) and the CS angle $\cos\theta^*$ (right).}
\end{figure}

In Fig.~\ref{fig:ResonantFid2} we show the rapidity distribution of the $\mu^+$ (left) and the $\cos\theta^*$ distribution (right).
The Collins--Soper~(CS) angle $\theta^{*}$~\cite{Collins:1977iv} is defined in
terms of LAB frame variables as
\begin{equation}
\label{eq:CS}
  \cos \theta^{*} = \frac{\ymumu}{|\ymumu|} \frac{2 (p_{\mu^{-}}^{+}p_{\mu^{+}}^{-} -p_{\mu^{-}}^{-}p_{\mu^{+}}^{+})} { \mmumu \sqrt{ \mmumu^{2} +  p_{T,\mu\mu}^{2}}}\,,
\end{equation}
with \mbox{$p^{\pm} = (p^{0} \pm p^{3})/\sqrt{2}$}.
The factor \mbox{${\ymumu}/{|\ymumu|}$} takes into account the fact that in
hadron collisions the quark and anti-quark directions are not known for
each collision. 
However, on average the quark carries more energy than the anti-quark as the latter must originate from the sea. 
Then, the produced dilepton system is, on average, boosted in the direction of the valence quark. Since the $Z$ boson rapidity is correlated with the direction of the valence quark, 
 it offers a sensible reference to define the scattering angle.

The radiative corrections to the rapidity distribution of the anti-muon
(Fig,~\ref{fig:ResonantFid2} (left))
have a rather uniform impact over the considered rapidity range. 
In the case of the $\cos\theta^*$ distribution, radiative corrections distort the distribution in different ways. NLO QCD corrections are larger at large $|\cos\theta^*|$, 
which is caused by the $qg$ channel opening up at NLO,
dominated by configurations where a valence quark at relatively large momentum fraction $x$ and a gluon at small $x$ enter the hard process.
By contrast, EW corrections distort the distribution in the opposite way, their impact being slightly larger and negative at large values of $|\cos\theta^*|$.
The impact of the mixed QCD--EW corrections is generally below the percent level, reaching the few percent level in the region of large $|\cos\theta^*|$.

\subsection{High invariant-mass region}
\label{sec:high}

Besides the resonant region, percent-level precision
can be envisaged also for data
taken in the region of high invariant masses. This requires a careful
assessment of the impact of EW radiative corrections and their interplay with
QCD effects. Following Ref.~\cite{CMS:2021ctt}, we consider proton--proton collisions at
\mbox{$\sqrt{s}=13\,$TeV} and impose the following cuts on the final-state leptons:
\begin{equation}\label{eq:HighMassFidCuts}
p_{T,\mu^{\pm}} > 53\,{\rm GeV}\,,\quad |y_{\mu}|<2.4\,,\quad \mmumu > 150\,{\rm GeV}\,.
\end{equation}
For this setup, a dynamical scale is a more appropriate choice. We use as central values of the renormalisation and factorisation scales the invariant mass of the dimuon system $\mmumu$. 
\begin{figure}[t]
  \centering
  \includegraphics[height=0.43\textheight]{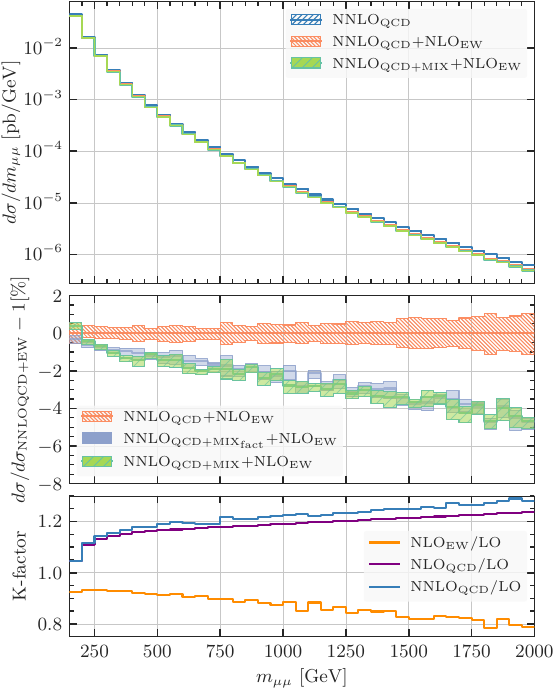}
  \caption{\label{fig:mll-CMSsetup} Dimuon invariant-mass distribution, in the acceptance setup defined in Section \ref{sec:high}. The structure of the panels and the coulour codes are the same as in Fig.~\ref{fig:ResonantFid1}. }
\end{figure}

In Fig.~\ref{fig:mll-CMSsetup} we show the invariant-mass distribution up to \mbox{$\mmumu=2\,$TeV}. Excluding the first bin close to the edge at
\mbox{$\mmumu=150\,$GeV}, imposed by the selection requirements on the leptons, the
effect of the mixed corrections is negative and increasing at higher invariant
masses, reaching $-5\%$ at \mbox{$\mmumu=2\,$TeV}.
Such values are comparable with or even larger than the statistical error expected at the end of the High-Luminosity (HL) phase of the LHC, with a total collected luminosity of $3\,$ab$^{-1}$. 
The inclusion of these corrections should help reducing the theoretical uncertainty on the partonic cross section of the DY process and, in turn, help constraining the proton PDFs, at large partonic $x$, in a global fit of collider data.\footnote{
The possibility that a PDF fit at large invariant masses inadvertently reabsorbs a New Physics signal has been discussed in Ref.~\cite{Hammou:2023heg}.} 
The precise knowledge of the higher-order corrections to this distribution is the mandatory starting point for an analysis which aims at the determination of the running of the $\overline{{\rm MS}}$ weak mixing angle~\cite{Amoroso:2023uux}.

We observe that the pattern of the mixed QCD--EW corrections is well approximated by
the factorised ansatz (see Eq.~(\ref{eq:fact})).
This implies that QCD and EW effects largely factorise in this region of high invariant masses, as also observed in Ref.~\cite{Buccioni:2022kgy}.

\begin{figure}[t]
  \centering
  \includegraphics[height=0.43\textheight]{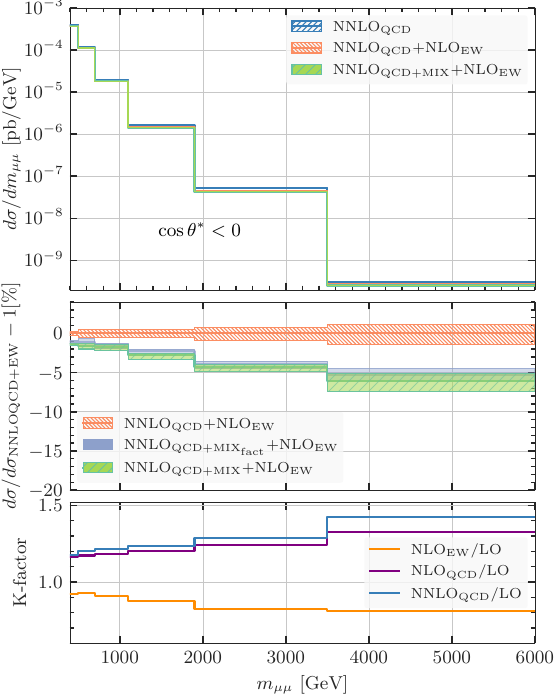}
  \hfill
  \includegraphics[height=0.43\textheight]{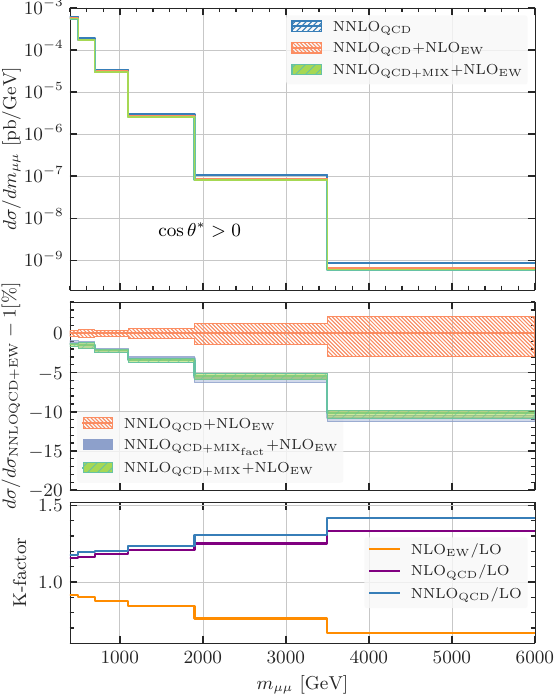}
  \caption{\label{fig:mll-CMSsetupBF} Dimuon invariant-mass distribution, in the acceptance setup defined in Section \ref{sec:high}, in the CMS bins~\cite{CMS:2021ctt}, in the backward (left) and forward (right) regions. The structure of the panels and the colour codes are the same as in Fig.~\ref{fig:ResonantFid1}.}
\end{figure}

In Fig.~\ref{fig:mll-CMSsetupBF}, we show the invariant-mass distributions in
the backward and forward regions up to \mbox{$\mmumu=6$\,TeV}. We see that the impact of mixed corrections increases as $\mmumu$ increases. The effect is slightly more pronounced in the forward region than in the backward region, following an analogous effect in the NLO EW corrections. As observed in Fig.~\ref{fig:mll-CMSsetup} the factorised ansatz works rather well.

\subsection{Forward--Backward asymmetry}
\label{sec:asym}

The Forward--Backward (FB) asymmetry $\afb$ is one of the main observables
for the determination of the leptonic effective weak mixing angle $\seffl$. 
At the LHC, $\afb$ is defined as the difference between the cross sections in the forward and backward directions
normalised to the total cross section, differentially with respect to the
invariant mass $m_{\ell\ell}$ of the dilepton system~\cite{Baur:1997wa,Baur:2001ze},
\begin{equation}
  \afb = \frac{F(\mll)-B(\mll)}{F(\mll)+B(\mll)}\,.
\end{equation}
The forward and backward cross sections are given by
\begin{equation}
  F(\mll) = \int_{0}^{1}d \cos\theta^{*}  \frac{d\sigma}{d \cos\theta^{*} d \mll}(\mll)\,, \quad 
  B(\mll) = \int_{-1}^{0}d \cos\theta^{*} \frac{d\sigma}{d \cos\theta^{*} d \mll}(\mll)\,,
\end{equation}
where $\cos\theta^*$ is defined in Eq.~(\ref{eq:CS}).

\begin{figure}
  \centering
  \includegraphics[width=\textwidth]{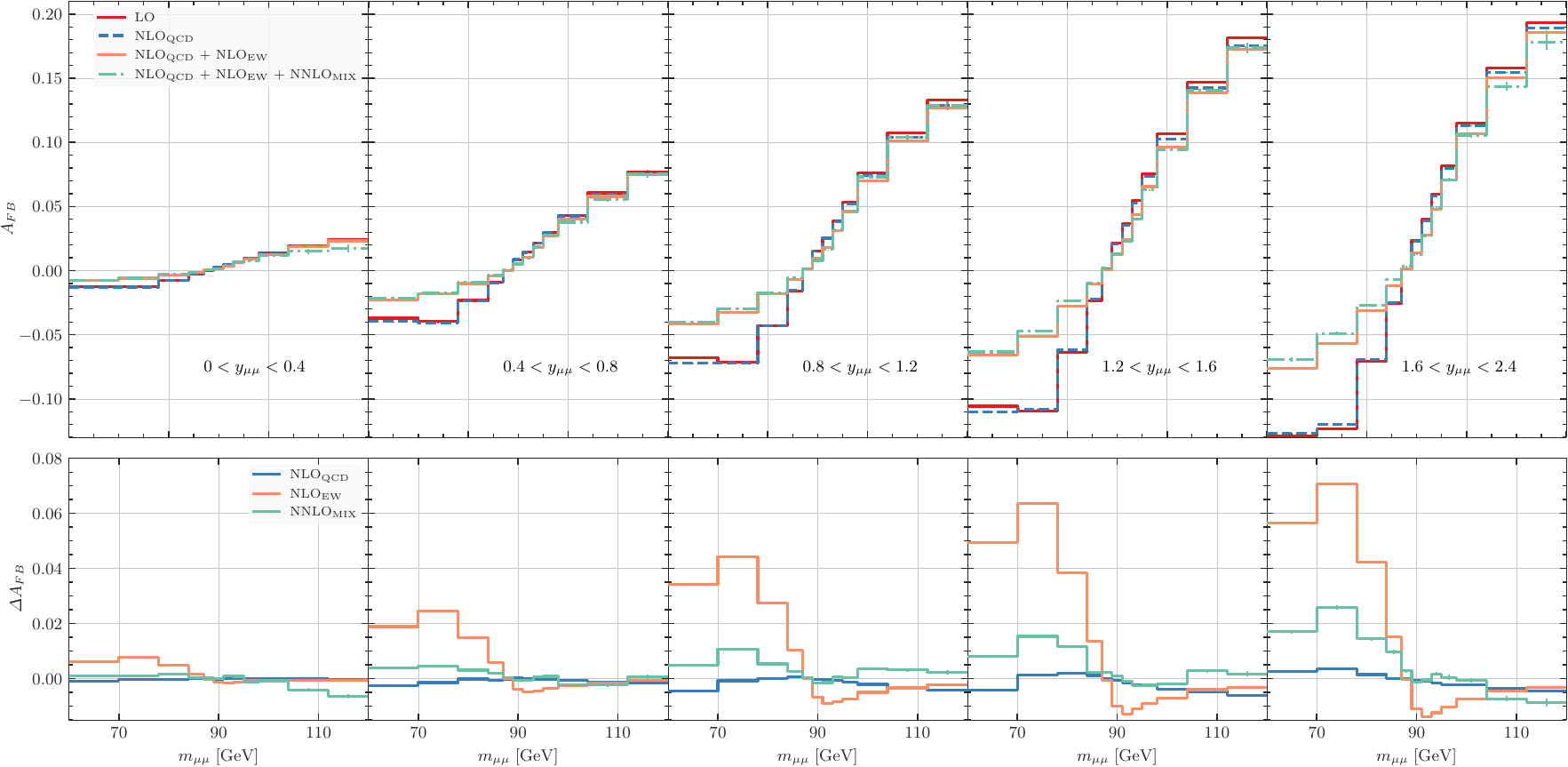}
  \caption{\label{fig:FB-asymmetry} Theoretical predictions for the FB
    asymmetry for different slices of rapidity of the dimuon final state at
    different perturbative orders. Shifts with respect to the LO prediction
    induced by separate sets of corrections are reported in the bottom panels.}
\end{figure}

The FB asymmetry is defined from the identification of the scattering angle of the negatively charged outgoing lepton, with respect to the direction of the incoming particle, in our case at LO a quark. 
In proton--proton collisions, we cannot prepare the partonic initial state or, in other words, we are not aware of the direction of the incoming quark: each final state receives, at LO, contributions from both quark--antiquark and antiquark--quark annihilation, listing the partons involved in the hard scattering process from the first and the second hadron, respectively.
If the partonic centre-of-mass system is at rest in the laboratory frame, \mbox{$y_{\ell\ell}=0$}, the invariance of the system under interchange of the incoming hadrons exactly cancels out all the parity-violating contributions to $\afb$.
At non-vanishing $y_{\ell\ell}$ values, the quark--antiquark subprocess (with the incoming quark oriented along the rapidity direction) prevails over the antiquark--quark process, because of the larger weight of the quark PDF with its valence content than that of the antiquark PDF; the unbalance avoids the cancellation of the parity-violating effects, and we observe a non-vanishing $\afb$. The slope of the asymmetry, in the $Z$ resonance region, is steeper  for larger $y_{\ell\ell}$ values, as illustrated by the different panels in Fig.~\ref{fig:FB-asymmetry}.

The experiments run at LEP/SLC~\cite{ALEPH:2005ab} have measured
the leptonic FB asymmetry at the $Z$ resonance with a precision of about $6\%$,
which in turn allowed a determination of $\seffl$ with a relative error of about $7\cdot 10^{-4}$.
Measurements of the FB asymmetry
have been performed by the CMS experiment at $8\,$TeV~\cite{CMS:2018ktx},
by the ATLAS experiment at $7\,$TeV~\cite{ATLAS:2015ihy} and 
by the LHCb experiment exploiting data at $7$ and $8\,$TeV~\cite{LHCb:2015jyu}. 
More recently, the CMS experiment has measured $\afb$ and extracted $\seffl$ from data collected during Run 2 at $13\,$TeV~\cite{CMS:2024aps}. 
These measurements have already reached a precision better than the percent level.

Given the foreseen increase in statistics in the HL-LHC phase, 
it is important to investigate the impact of higher-order QCD, EW, and mixed QCD--EW radiative corrections in the theory predictions of $\afb$.
These effects have been recently studied in the pole
approximation~\cite{Dittmaier:2024row} in proton--proton collisions at $13\,$TeV.
In the following, we present a study on the impact of the exact mixed QCD--EW corrections to the FB asymmetry,
for dimuon production in proton--proton collisions at $8\,$TeV.
We adopt the setup used by the CMS collaboration for the extraction of $\seffl$~\cite{CMS:2018ktx}:
\begin{equation}\label{eq:Re-staggered}
  p_{T,1}> 25\,{\rm GeV}\,,\quad p_{T,2}> 15\,{\rm GeV}\,, \quad |y_{\mu}|<2.4\,, \quad 60 < \mmumu < 120\,{\rm GeV}\,,
\end{equation}
where $p_{T,1}$\,($p_{T,2}$) is the transverse momentum of the hardest (second hardest) lepton. 
We consider five intervals in the absolute rapidity of the dimuon final
state: \mbox{$0<|\yll|<0.4$}, \mbox{$0.4<|\yll|<0.8$}, \mbox{$0.8<|\yll|<1.2$},
\mbox{$1.2<|\yll|<1.6$} and \mbox{$1.6<|\yll|<2.4$}. 
In order to better quantify the impact of the higher-order corrections, we consider the shifts with respect to the LO prediction for the FB asymmetry,
\begin{align}
  \Delta A_{FB}^{\rm X}(m_{\ell\ell}) = A_{FB}^{\rm X}(m_{\ell\ell}) -A_{FB}^{\rm LO}(m_{\ell\ell})\;, 
\end{align}
where ${\rm X=NLO_{QCD}}$, ${\rm NLO_{EW}}$ or ${\rm NNLO_{MIX}}$\footnote{Here ${\rm NNLO_{MIX}}$ includes only the mixed QCD--EW corrections on top of the LO prediction.}. 
In Fig.~\ref{fig:FB-asymmetry}, we show results for $\afb$ with an increasing radiative content. 
The shifts compared to the LO asymmetry are shown in the bottom panels of Fig.~\ref{fig:FB-asymmetry}. 

We first observe that the NLO QCD corrections have a tiny impact throughout the whole invariant-mass range: the strong interaction is invariant under parity, and QCD corrections largely cancel in the definition of $\afb$.
The NLO EW corrections have a large impact, through two distinct mechanisms: the large parity-even QED correction contributes mostly due to final-state radiation, which is particularly important below the $Z$ resonance; the purely weak correction contributes instead to the parity-violating contribution. These two effects combined yield the largest shift of $\afb$.
The mixed QCD--EW corrections inherit both features from the NLO EW terms, with increasing impact at larger $\yll$ values; they are larger in size than the purely QCD corrections.

The effective weak mixing angle $\seffl$ can be determined from the study of $\afb$~\cite{Chiesa:2019nqb}.
In the $Z$-resonance region $\afb$ is generated by the product of vector and axial-vector couplings of the $Z$ boson to fermions, in the squared $Z$-exchange diagrams, yielding in turn a sensitivity to the value of the weak mixing angle, present in the vector coupling. Away from the resonance, the squared $Z$-exchange diagrams are kinematically suppressed, $\afb$ is larger than at the resonance and driven by the product of the photon vector coupling with the axial coupling of the $Z$ boson. This second coupling combination has no dependence on the weak mixing angle. 

The $\seffl$ determination cannot be based on the analysis of one single bin of the $\afb$ invariant-mass distribution, i.e.\ the $Z$-resonance bin, because the result would be swamped by PDF uncertainties. The latter can be constrained~\cite{CMS:2024aps}, adopting a template fit procedure, which typically requires the analysis of $\afb$ in the mass window $[70,110]\,$GeV. It is thus necessary to have control over the radiative corrections in this very region.

\begin{figure}
    \centering
    \includegraphics[width=\textwidth]{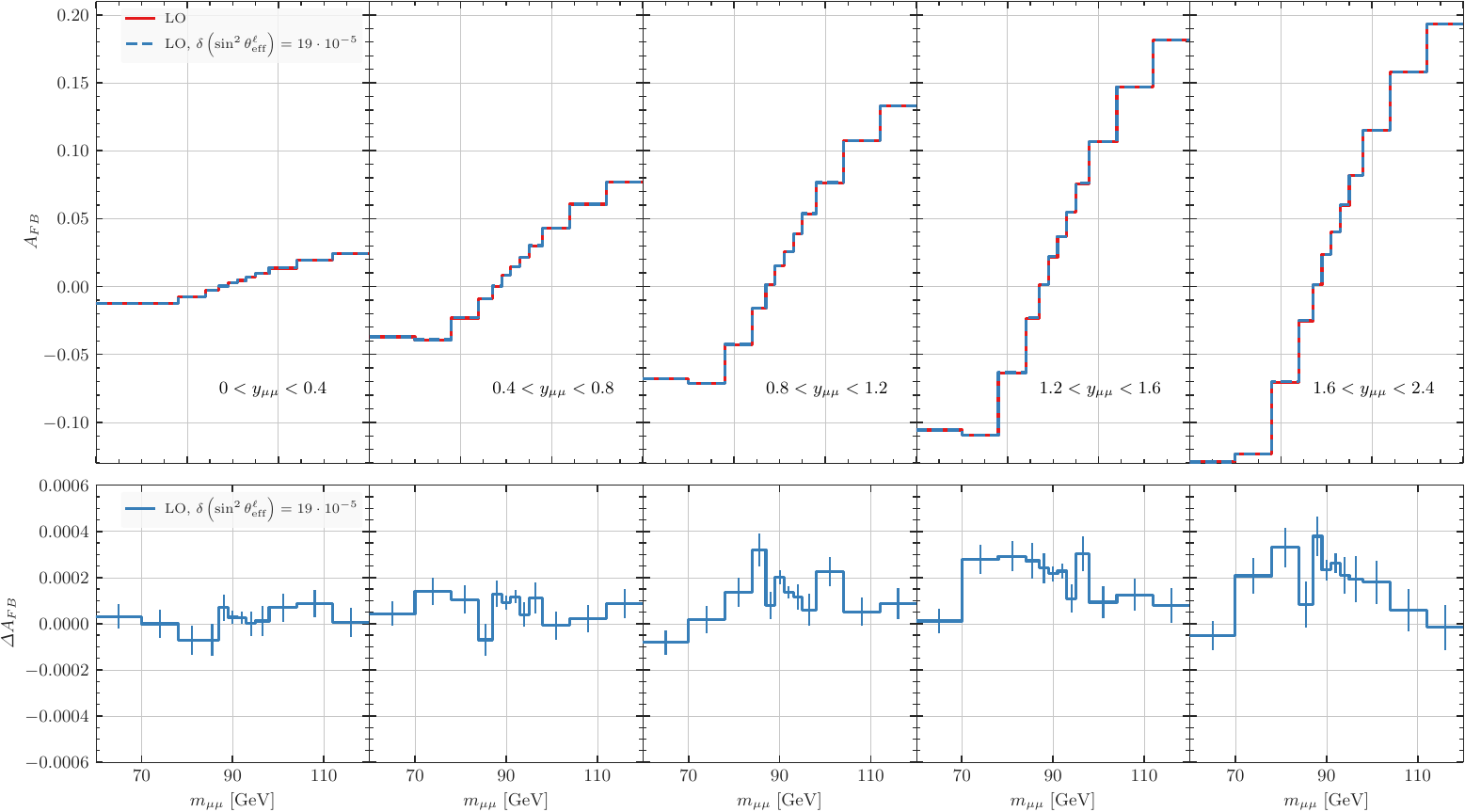}
    \caption{The variation of $\afb$ for a change of the weak mixing angle value by \mbox{$19\cdot 10^{-5}$}.}
    \label{fig:FB-sensitivity}
\end{figure}

In order to illustrate the sensitivity of $\afb$ to $\seffl$, in Fig.~\ref{fig:FB-sensitivity} we study the variation of the asymmetry induced by a change\footnote{Since our calculation is performed with $(G_\mu, m_W, m_Z, m_H)$ as input parameters of the EW Lagrangian, we induce a variation of the weak mixing angle by a change of the value of the $W$-boson mass of 10 MeV.} of the weak mixing angle by \mbox{$1.9\cdot 10^{-4}$}, a value slightly larger than the current experimental error on $\seffl$ \cite{ParticleDataGroup:2024cfk}.
For the sensitivity estimate, it is sufficient to consider a variation of the LO $\afb$ asymmetry, which we present in Fig.~\ref{fig:FB-sensitivity}. The extraction of $\seffl$ from a fit to the experimental data must be performed using the most precise theoretical templates: missing higher orders could affect the final result inducing a spurious shift of the central value. The size of the latter can be estimated evaluating the impact of the higher order corrections on $\afb$. 

Comparing Fig.~\ref{fig:FB-asymmetry} and Fig.~\ref{fig:FB-sensitivity}, we clearly see that the impact of the mixed QCD--EW corrections is large.
Part of the effect is known to be due to
QED final-state radiation~\cite{Dittmaier:2024row}, which is typically accounted for in the experimental analyses
using Monte Carlo parton showers. Nonetheless, when comparing
Fig.~\ref{fig:FB-asymmetry} and Fig.~\ref{fig:FB-sensitivity}, it is evident
that a thorough evaluation of the computed corrections will be essential for
accurate future determinations of the effective electroweak mixing angle
$\seffl$.

\section{Comparison with massless calculation}
\label{sec:massless}

In this Section we present a comparison of our results to those of Ref.~\cite{Buccioni:2022kgy}.
To this purpose, the EW scheme and input parameters are tuned to those of Ref.~\cite{Buccioni:2022kgy}, as far as possible.
In Ref.~\cite{Buccioni:2022kgy} the selection cuts are
\begin{equation}
\label{eq:pcuts}
  m_{\ell\ell}>200\,{\rm GeV}, \quad p_{T,\ell^{\pm}}> 30\,{\rm GeV}, \quad \sqrt{p_{T,\ell^{+}}p_{T,\ell^{-}}}>35\,{\rm GeV}, \quad |y_{\ell^{\pm}}|< 2.5\;,
\end{equation}
and leptons are treated as massless. 

It is well known that
fiducial cuts applied to {\it bare} massless
leptons are, in general, not collinear safe with respect to the
emission of a collinear photon off a final-state lepton. The
definition of physical observables requires therefore the introduction of
a recombination procedure of photons to a close-by lepton\footnote{In Ref.~\cite{Buccioni:2022kgy} a lepton and a photon are recombined if their separation in rapidity and azimuth $R_{\ell\gamma}=\sqrt{\Delta^2 y_{\ell\gamma}+\Delta\phi_{\ell\gamma}^2}$ is smaller than a parameter $R_{\rm cut}=0.1$}. This
prevented the authors of Ref.~\cite{Buccioni:2022kgy} to directly
compare their results, valid for {\it dressed} or {\it calo}
leptons, to the results of Ref.~\cite{Bonciani:2021zzf}, obtained for the case of bare massive muons.

Conversely, our framework, which retains the dependence on the mass of
the final-state leptons, allows us to perform calculations for both
bare and dressed leptons. This can be easily understood if we take
seriously the fact that the mass of the lepton acts as the regulator
of the final-state collinear singularity associated with the emission of
a photon off a lepton. Indeed, the resulting dead-cone effect leads to
a smooth suppression of the collinear photon radiation at the parton
level. In combination with the \mbox{ $q_T/m_{\ell\ell}>\rcut$} requirement, applied at the parton/bare level, this makes
the real-emission cross section finite. We can then apply the
recombination procedure to each parton level event, analogously to what happens for
jet cross sections computed with a local subtraction method.
We also stress that the mass of the lepton should not be regarded as an additional cut-off parameter in our
formalism (besides the slicing parameter $\rcut$). Our subtraction formula in Eq.~(\ref{eq:master}) indeed retains the exact
dependence on the mass of the lepton, without introducing further
approximations.

To summarise, we can compare with the results of
Ref.~\cite{Buccioni:2022kgy} for dressed leptons within our framework
by
repeating the calculation for decreasing lepton masses and eventually taking
  the limit \mbox{$m_\ell\to 0$}.
In practice, we can perform the calculation at a sufficiently small value of the lepton mass, which
can be even larger than the physical mass of the considered leptons, as long as
power corrections can be safely neglected. Our comparison with the results for dressed electrons of Ref.~\cite{Buccioni:2022kgy} is carried out using a reference lepton mass
\mbox{$m_\ell = m_\mu = 105.658369\,$MeV.}

The above procedure is consistent with the fact that we are performing an
on-shell renormalisation of the electric charge. Instead, one can consider a
running coupling constant as, for example, in the $\overline{\rm MS}$ scheme
with only massless fermions (quarks and leptons) actively running in the beta
function, while the wave functions of the external massive leptons are renormalised
on-shell (decoupling scheme). In this situation, it is well known that an extra
correction contribution, proportional to $\ln\mu_{R}/m_{\ell}$, must be
added in passing from the massive to the massless calculation in order to compensate
for the different number of active flavours in the beta function~\cite{Cacciari:1998it}.

There is, however, a caveat related to LO photon-induced processes, namely the
\mbox{$\gamma\gamma \to \ell^+ \ell^-$} channel. In this case, the subtraction of the
photonic vacuum polarisation insertions is usually dealt with relying on
dimensional regularisation for the light flavours, whereas, for massive fermions, the
mass is used as the physical regulator. This is consistent with the reabsorption of
initial-state collinear singularities into the renormalisation of the parton
distribution functions when applying collinear factorisation, with the light
flavours treated as active in the $\overline{\rm MS}$ scheme, whereas the massive flavours
are only generated radiatively from scales above the pair production threshold.
Therefore, in this case, the procedure stated above must be supplemented by
adding a contribution which compensates for the different treatment in the
collinear factorisation. In other words, the photon distribution receives an ${\cal O}(\alpha)$ correction due to the change of scheme~\cite{Collins:1986mp}, and the cross section in the $\gamma\gamma$ channel is modified as
\begin{equation}
d\sigma^{(0,1)}_{\gamma\gamma,m_{\ell}=0} = d\sigma^{(0,1)}_{\gamma\gamma,m_{\ell}} - \alpha \frac{2e^{2}_{\ell}}{3 \pi} \ln \frac{m_{\ell}^{2}}{\mu_{F}^{2}} d\sigma^{(0,0)}_{\gamma\gamma,m_{\ell}}\,,
\end{equation}
which is obtained through the abelianisation of the analogous formula for the case of the $gg$ channel given in Ref.~\cite{Cacciari:1998it}.

In Tab.~\ref{tab:comparison-MK}, we show our results for different orders in the
different partonic channels. Our predictions for the NLO EW corrections are in
very good agreement with those of Ref.~\cite{Buccioni:2022kgy}. The small
difference for the quark--antiquark ($q\bar{q}$) channel~\footnote{For
  simplicity, the quark--antiquark channel is assumed to include all possible scattering
  processes involving quark–antiquark, quark–quark, or antiquark–antiquark initial states of any flavour.} is related to the fact we are using a finite
lepton mass. Nonetheless, such a difference is well below the $0.01\%$ level for
fiducial cross sections, confirming that the power corrections in the lepton
mass are indeed negligible for the chosen value.
\begin{table}
  \centering
  \small
\renewcommand{\arraystretch}{1.4}
  \centering
  \begin{tabular}{|c|c|c|c|c|c|}
    \hline
    $\sigma$ [pb] & $\sigma_{\rm LO}$  & $\sigma^{(1,0)}$  & $\sigma^{(0,1)}$  & $\sigma^{(2,0)}$ & $\sigma^{(1,1)}$ \\
    \hline
    \hline
    $q{\bar q}$ & $1561.52(5)$ & $340.3(3)$ & $-49.77(5)$ & $44.6(4)$ &$-15.7(7)$\\
    \hline
    $qg$ & --- & $0.0601(3)$ & --- & $-32.7(2)$ & $2.15(7)$ \\
    \hline
    $q\gamma$ & --- & --- & $-0.30(2)$ & --- & $-0.231(9)$\\
    \hline
    $g\gamma$ & --- & --- & --- & -- & $0.266(1)$\\
    \hline
    $g g$ & --- & --- & --- & $2.02(6)$ & --- \\
     \hline
    $\gamma \gamma$ & $59.645(6)$ & --- & $3.174(9)$ & --- & --- \\
    \hline
  \end{tabular}
  \caption{\label{tab:comparison-MK}
    The different perturbative contributions to the fiducial cross section in the various partonic channels, in the setup of Ref.~\cite{Buccioni:2022kgy}.}
\renewcommand{\arraystretch}{1.0}  
\end{table}

Turning to higher orders, we observe a convincing agreement for the NNLO QCD
corrections in all considered partonic channels with the corresponding results
of Ref.~\cite{Buccioni:2022kgy}. The choice of {\it product cuts}
\cite{Salam:2021tbm} in Eq.~(\ref{eq:pcuts}) alleviates the problem of linear
power corrections in the $q_T$ subtraction formalism, leading to a better
convergence of the method. Concerning the mixed QCD--EW corrections, we find
agreement within uncertainties in the quark--antiquark
channel, which contains the genuine two-loop virtual contribution, and in the
gluon--photon channel. Our result for the correction in the quark--photon
channel is about $10\%$ larger than the corresponding result of
Ref.~\cite{Buccioni:2022kgy}.
We note that Ref.~\cite{Buccioni:2022kgy} quotes an overall $1\%$ uncertainty on
the mixed QCD--EW corrections, but does not provide the numerical accuracy of the
separate partonic contributions, so we are unable to assess the significance of
this small discrepancy. In the quark--gluon channel, however, the discrepancy is
definitely significant, our result being about a factor of two larger than the
corresponding result of Ref.~\cite{Buccioni:2022kgy}. Despite our efforts and
two completely independent numerical implementations, we were unable to explain
this difference. After extensive cross checks, the authors of Ref.~\cite{Buccioni:2022kgy} informed us that
they found a bug in their implementation. After fixing this bug their result in the quark--gluon is in perfect agreement with ours. More detailed checks, also at the differential level, are needed for a thorough comparison of the two calculations.

\begin{figure}
  \centering
  \includegraphics[height=0.29\textheight]{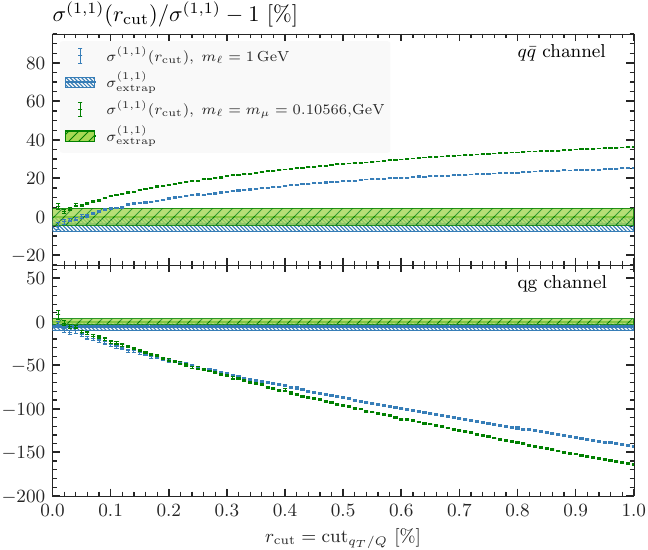}
  \hfill 
  \includegraphics[height=0.29\textheight]{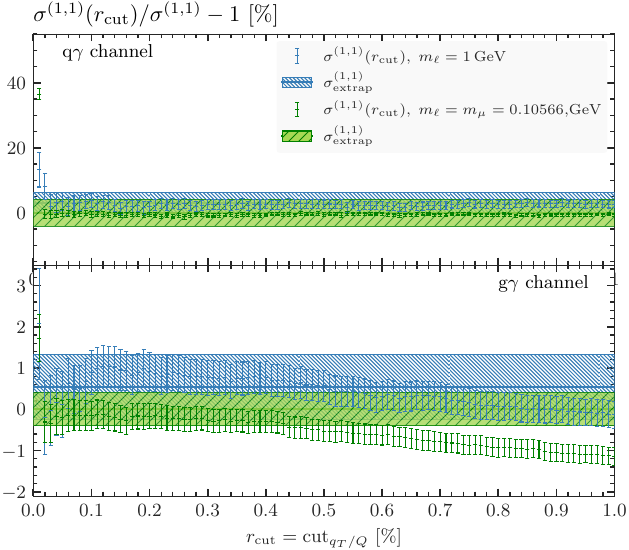}
  \caption{\label{fig:RcutDepCMP} Mixed QCD--EW corrections $\sigma^{(1,1)}$ as functions of the parameter $\rcut$ in the quark--antiquark (top left), quark--gluon (bottom left), quark--photon (top right) and gluon--photon (bottom right) channels for two different values of the final-state lepton mass, \mbox{$m_{\ell}=1\,$GeV} and \mbox{$m_{\ell}=m_{\mu}$}.
    Our results are normalised to the \mbox{$m_{\ell}=m_{\mu}$} result in the $\rcut \to 0$ limit.}
\end{figure}

In Fig.~\ref{fig:RcutDepCMP}, we study the
dependence on the slicing parameter $\rcut$ for the nominal lepton mass
\mbox{$m_\ell=m_\mu$} and for a larger mass value, \mbox{$m_\ell = 1\,$GeV}. The results are
consistent with the assumption that the residual dependence on the lepton mass,
after the cancellation of large logarithms between real and virtual
contributions, is indeed power suppressed. We observe that the power corrections of the lepton mass have a marginal
impact on the final result. For comparison, the effect of increasing the lepton mass to \mbox{$m_\ell = 1\,$GeV} on NLO EW corrections is to reduce them by about $3\%$. Our results provide a strong
confirmation that the cancellation of all IR divergencies in our method does
take place correctly, thereby providing a stringent check of the whole
calculation.

\begin{figure}
  \centering
  \includegraphics[height=0.43\textheight]{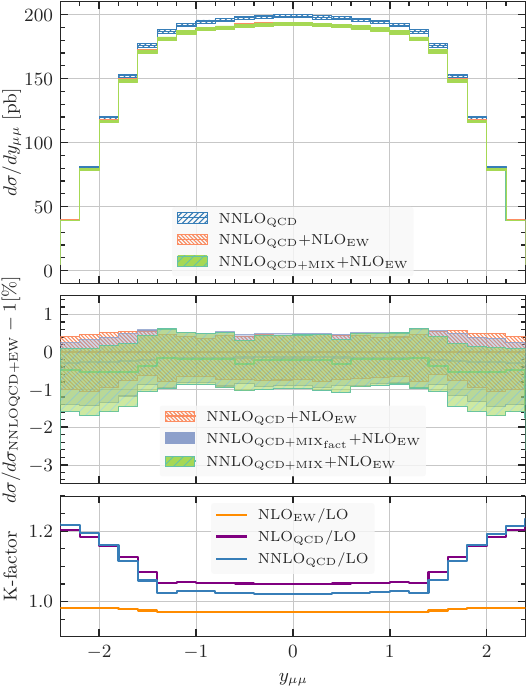}
  \hfill
  \includegraphics[height=0.43\textheight]{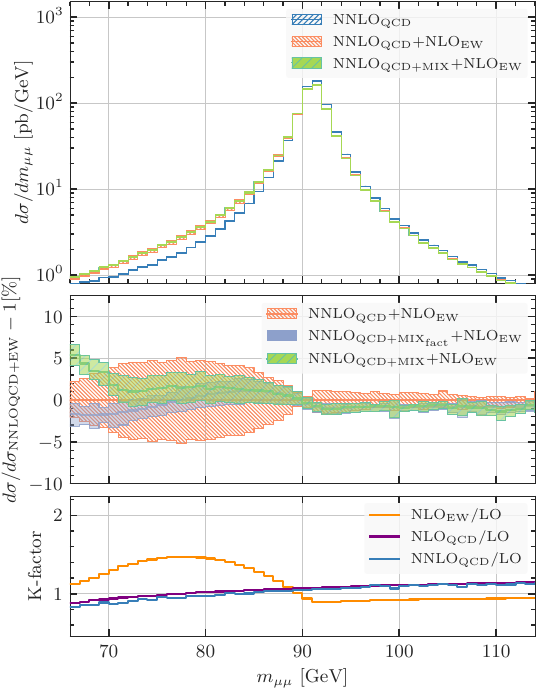}
  \caption{\label{fig:ResonantFid1-dressed} As in Fig.~\ref{fig:ResonantFid1} but for dressed leptons.
}
\end{figure}

We conclude this section by presenting further results for dressed leptons, which correspond to the setup of Sect.~\ref{sec:reso}. In Table~\ref{tab:res-dressed}, we show the corresponding fiducial cross
  sections. Figure~\ref{fig:ResonantFid1-dressed} displays the rapidity spectrum
  and the invariant mass distribution of the dilepton system. As discussed
  above, we use a fixed lepton mass, setting $m_{l}=m_{\mu}$, with negligible
  power corrections. As expected, the effect of QED final-state radiation
  is reduced when considering dressed leptons, leading to a decrease in
  radiative corrections in the NLO EW and mixed QCD-EW contributions. This
  effect is more pronounced in the invariant mass distribution, where the
  radiative tail below the resonance peak is significantly suppressed.

\begin{table}[h!]
\small
\renewcommand{\arraystretch}{1.4}
  \centering
    \begin{tabular}{l | c c c }
     & $\sigma$ [pb]  &  $\sigma^{(i,j)}$ [pb] &  $\sigma^{(i,j)}/\sigma_{\rm LO}$
     \\ \midrule
LO & $735.73(3)^{+12.7\%}_{ -13.6\%}$ & $-$ & $-$
       \\ [1.5ex]
NLO$_{\rm QCD}$ & $790.23(4)^{+2.7\%}_{ -4.4\%}$ & $\sigma^{(1,0)}= +54.50(5)$ & $+7.4$\%\\[1.5ex]
NNLO$_{\rm QCD}$ & $774.4(7)^{+0.7\%}_{-0.7\%}$ & $\sigma^{(2,0)}=-15.8(5)\phantom{0}$ & $-2.2$\%\\ [1.5ex]
NLO$_{\rm EW}$ & $715.94(8)^{+12.7}_{ -13.6\%}$ & $-19.7(1)$ & $-2.7$\%\\
NNLO$_{\rm QCD}$+NLO$_{\rm EW}$ & $754.6(5)^{+0.4\%}_{ -0.6\%}$ & $-$ & $-$\\
NNLO$_{\rm QCD+MIX}$+NLO$_{\rm EW}$ & $752.4(5)^{+0.7\%}_{ -0.6\%}$ & $-2.1(1)$ & $-0.3$\% \\
\bottomrule
  \end{tabular}
\caption{\label{tab:res-dressed} As in Tab.~\ref{tab:res} but for dressed leptons.}
\end{table}

\section{Summary}
\label{sec:summa}

In this paper we have reported on the complete computation of the mixed QCD--EW
corrections to the neutral-current Drell--Yan process. Our calculation does not
rely on any approximation and holds in the entire range of dilepton invariant
masses. We have presented selected phenomenological results for several
kinematical distributions in the case of bare muons both in the resonant and in
the high invariant-mass region.

The impact of the newly computed corrections is typically at the few per mille
level for the fiducial cross section, but can reach the percent level in some
kinematical regions, and for specific kinematical distributions like the
invariant mass of the dilepton pair for bare muons, thereby confirming their
relevance for precision studies of the Drell--Yan process. The lepton-pair rapidity distribution receives mixed QCD--EW corrections ranging from few per mille in the central region up to 1\%  at large rapidities, with a possible impact on precision fits of the proton PDFs.

We have also considered the forward--backward asymmetry $\afb$, which is one of
the main observables for the determination of the leptonic effective weak mixing
angle at hadron colliders and is also potentially sensitive to new heavy resonances~\cite{Ball:2022qtp}. We studied the impact of the mixed QCD--EW corrections
on $\afb$ by using the same setup adopted in the CMS measurement at
$\sqrt{s}=8\,$TeV~\cite{CMS:2018ktx}. We have shown that the impact of mixed QCD--EW corrections is significant. Since part of the effect is known to be due to QED final-state radiation
and is therefore accounted for in the experimental analyses, a precise estimate of the impact of the computed corrections in $\seffl$ determinations will require more detailed studies.

We have finally applied our calculation to the case of dressed leptons. We have shown that in this case our numerical computation is sufficiently stable to be extrapolated to the massless limit. This is definitely non-trivial, given that the calculation is carried out by using a slicing method~\cite{Catani:2007vq} in which final-state collinear singularities are regulated just with a finite lepton mass. We were able to compare
our results for dressed leptons to those of Ref.~\cite{Buccioni:2022kgy}. We found a relatively good agreement between the corresponding results for the fiducial cross section except for the quark--gluon channel, for which our results disagree with those of Ref.~\cite{Buccioni:2022kgy}. This disagreement disappears after the correction of a bug in the implementation of Ref.~\cite{Buccioni:2022kgy}.
Further cross-checks with the authors of Ref.~\cite{Buccioni:2022kgy} are ongoing.

\noindent {\bf Acknowledgements}

We are grateful to Federico Buccioni for his efforts in cross checking with the results of Ref.~\cite{Buccioni:2022kgy}.
We thank Francesco Tramontano for contributions and insights at the early stage of this work. This work is supported in part by the Swiss National Science Foundation (SNF)
under contracts 200020$\_$188464 and 200020$\_$219367. TA is a Research Fellow of the Fonds de la Recherche Scientifique – FNRS. The work of LB is funded
by the European Union (ERC, grant agreement No. 101044599, JANUS). The work
of SD is funded by the European Union (ERC, Grant
Agreement No. 101078449, MultiScaleAmp). Views and
opinions expressed are however those of the authors only and do not necessarily
reflect those of the European Union or the European Research Council Executive
Agency. Neither the European Union nor the granting authority can be held
responsible for them.

\noindent

\bibliography{biblio}

\providecommand{\href}[2]{#2}\begingroup\raggedright\begin{thebibliography}{100}

\bibitem{Drell:1970wh}
S.~Drell and T.-M. Yan, {\it {Massive Lepton Pair Production in Hadron-Hadron
  Collisions at High-Energies}},  {\em Phys. Rev. Lett.} {\bf 25} (1970)
  316--320. [Erratum: Phys.Rev.Lett. 25, 902 (1970)].

\bibitem{Amoroso:2022eow}
S.~Amoroso et~al., {\it {Snowmass 2021 Whitepaper: Proton Structure at the
  Precision Frontier}},  {\em Acta Phys. Polon. B} {\bf 53} (2022), no.~12
  12--A1, [\href{http://arxiv.org/abs/2203.13923}{{\tt arXiv:2203.13923}}].

\bibitem{ATLAS:2017rzl}
{\bf ATLAS} Collaboration, M.~Aaboud et~al., {\it {Measurement of the $W$-boson
  mass in pp collisions at $\sqrt{s}=7$ TeV with the ATLAS detector}},  {\em
  Eur. Phys. J. C} {\bf 78} (2018), no.~2 110,
  [\href{http://arxiv.org/abs/1701.07240}{{\tt arXiv:1701.07240}}]. [Erratum:
  Eur.Phys.J.C 78, 898 (2018)].

\bibitem{LHCb:2021bjt}
{\bf LHCb} Collaboration, R.~Aaij et~al., {\it {Measurement of the W boson
  mass}},  {\em JHEP} {\bf 01} (2022) 036,
  [\href{http://arxiv.org/abs/2109.01113}{{\tt arXiv:2109.01113}}].

\bibitem{ATLAS:2024erm}
{\bf ATLAS} Collaboration, G.~Aad et~al., {\it {Measurement of the W-boson mass
  and width with the ATLAS detector using proton-proton collisions at
  $\sqrt{s}$ = 7 TeV}},  \href{http://arxiv.org/abs/2403.15085}{{\tt
  arXiv:2403.15085}}.

\bibitem{CMS:2024nau}
{\bf CMS} Collaboration, {\it {Measurement of the W boson mass in proton-proton
  collisions at $\sqrt s$ = 13 TeV}}, .

\bibitem{Aaltonen:2018dxj}
{\bf CDF, D0} Collaboration, T.~A. Aaltonen et~al., {\it {Tevatron Run II
  combination of the effective leptonic electroweak mixing angle}},  {\em Phys.
  Rev. D} {\bf 97} (2018), no.~11 112007,
  [\href{http://arxiv.org/abs/1801.06283}{{\tt arXiv:1801.06283}}].

\bibitem{LHCb:2015jyu}
{\bf LHCb} Collaboration, R.~Aaij et~al., {\it {Measurement of the
  forward-backward asymmetry in $Z/\gamma^{\ast} \rightarrow \mu^{+}\mu^{-}$
  decays and determination of the effective weak mixing angle}},  {\em JHEP}
  {\bf 11} (2015) 190, [\href{http://arxiv.org/abs/1509.07645}{{\tt
  arXiv:1509.07645}}].

\bibitem{ATLAS:2018gqq}
{\bf ATLAS} Collaboration, {\it {Measurement of the effective leptonic weak
  mixing angle using electron and muon pairs from $Z$-boson decay in the ATLAS
  experiment at $\sqrt s = 8$ TeV}},  {\em ATLAS-CONF-2018-037} (7, 2018).

\bibitem{CMS:2024aps}
{\bf CMS} Collaboration, {\it {Measurement of the Drell-Yan forward-backward
  asymmetry and of the effective leptonic weak mixing angle using proton-proton
  collisions at 13 TeV}}, .

\bibitem{Ball:2018iqk}
{\bf NNPDF} Collaboration, R.~D. Ball, S.~Carrazza, L.~Del~Debbio, S.~Forte,
  Z.~Kassabov, J.~Rojo, E.~Slade, and M.~Ubiali, {\it {Precision determination
  of the strong coupling constant within a global PDF analysis}},  {\em Eur.
  Phys. J. C} {\bf 78} (2018), no.~5 408,
  [\href{http://arxiv.org/abs/1802.03398}{{\tt arXiv:1802.03398}}].

\bibitem{ATLAS:2023lhg}
{\bf ATLAS} Collaboration, G.~Aad et~al., {\it {A precise determination of the
  strong-coupling constant from the recoil of $Z$ bosons with the ATLAS
  experiment at $\sqrt{s} = 8$ TeV}},
  \href{http://arxiv.org/abs/2309.12986}{{\tt arXiv:2309.12986}}.

\bibitem{Altarelli:1979ub}
G.~Altarelli, R.~Ellis, and G.~Martinelli, {\it {Large Perturbative Corrections
  to the Drell-Yan Process in QCD}},  {\em Nucl. Phys. B} {\bf 157} (1979)
  461--497.

\bibitem{Hamberg:1990np}
R.~Hamberg, W.~van Neerven, and T.~Matsuura, {\it {A complete calculation of
  the order $\alpha_s^{2}$ correction to the Drell-Yan $K$ factor}},  {\em
  Nucl. Phys. B} {\bf 359} (1991) 343--405. [Erratum: Nucl.Phys.B 644, 403--404
  (2002)].

\bibitem{Harlander:2002wh}
R.~V. Harlander and W.~B. Kilgore, {\it {Next-to-next-to-leading order Higgs
  production at hadron colliders}},  {\em Phys. Rev. Lett.} {\bf 88} (2002)
  201801, [\href{http://arxiv.org/abs/hep-ph/0201206}{{\tt hep-ph/0201206}}].

\bibitem{Anastasiou:2003yy}
C.~Anastasiou, L.~J. Dixon, K.~Melnikov, and F.~Petriello, {\it {Dilepton
  rapidity distribution in the Drell-Yan process at NNLO in QCD}},  {\em Phys.
  Rev. Lett.} {\bf 91} (2003) 182002,
  [\href{http://arxiv.org/abs/hep-ph/0306192}{{\tt hep-ph/0306192}}].

\bibitem{Anastasiou:2003ds}
C.~Anastasiou, L.~J. Dixon, K.~Melnikov, and F.~Petriello, {\it {High precision
  QCD at hadron colliders: Electroweak gauge boson rapidity distributions at
  NNLO}},  {\em Phys. Rev. D} {\bf 69} (2004) 094008,
  [\href{http://arxiv.org/abs/hep-ph/0312266}{{\tt hep-ph/0312266}}].

\bibitem{Melnikov:2006kv}
K.~Melnikov and F.~Petriello, {\it {Electroweak gauge boson production at
  hadron colliders through $\mathcal{O}(\alpha_s^2)$}},  {\em Phys. Rev. D}
  {\bf 74} (2006) 114017, [\href{http://arxiv.org/abs/hep-ph/0609070}{{\tt
  hep-ph/0609070}}].

\bibitem{Catani:2009sm}
S.~Catani, L.~Cieri, G.~Ferrera, D.~de~Florian, and M.~Grazzini, {\it {Vector
  boson production at hadron colliders: a fully exclusive QCD calculation at
  NNLO}},  {\em Phys. Rev. Lett.} {\bf 103} (2009) 082001,
  [\href{http://arxiv.org/abs/0903.2120}{{\tt arXiv:0903.2120}}].

\bibitem{Catani:2010en}
S.~Catani, G.~Ferrera, and M.~Grazzini, {\it {W Boson Production at Hadron
  Colliders: The Lepton Charge Asymmetry in NNLO QCD}},  {\em JHEP} {\bf 05}
  (2010) 006, [\href{http://arxiv.org/abs/1002.3115}{{\tt arXiv:1002.3115}}].

\bibitem{Dittmaier:2001ay}
S.~Dittmaier and M.~Kr\"amer, {\it {Electroweak radiative corrections to W
  boson production at hadron colliders}},  {\em Phys. Rev. D} {\bf 65} (2002)
  073007, [\href{http://arxiv.org/abs/hep-ph/0109062}{{\tt hep-ph/0109062}}].

\bibitem{Baur:2004ig}
U.~Baur and D.~Wackeroth, {\it {Electroweak radiative corrections to $p \bar{p}
  \to W^\pm \to \ell^\pm \nu$ beyond the pole approximation}},  {\em Phys. Rev.
  D} {\bf 70} (2004) 073015, [\href{http://arxiv.org/abs/hep-ph/0405191}{{\tt
  hep-ph/0405191}}].

\bibitem{Zykunov:2006yb}
V.~Zykunov, {\it {Radiative corrections to the Drell-Yan process at large
  dilepton invariant masses}},  {\em Phys. Atom. Nucl.} {\bf 69} (2006) 1522.

\bibitem{Arbuzov:2005dd}
A.~Arbuzov, D.~Bardin, S.~Bondarenko, P.~Christova, L.~Kalinovskaya, G.~Nanava,
  and R.~Sadykov, {\it {One-loop corrections to the Drell-Yan process in SANC.
  I. The Charged current case}},  {\em Eur. Phys. J. C} {\bf 46} (2006)
  407--412, [\href{http://arxiv.org/abs/hep-ph/0506110}{{\tt hep-ph/0506110}}].
  [Erratum: Eur.Phys.J.C 50, 505 (2007)].

\bibitem{CarloniCalame:2006zq}
C.~Carloni~Calame, G.~Montagna, O.~Nicrosini, and A.~Vicini, {\it {Precision
  electroweak calculation of the charged current Drell-Yan process}},  {\em
  JHEP} {\bf 12} (2006) 016, [\href{http://arxiv.org/abs/hep-ph/0609170}{{\tt
  hep-ph/0609170}}].

\bibitem{Baur:2001ze}
U.~Baur, O.~Brein, W.~Hollik, C.~Schappacher, and D.~Wackeroth, {\it
  {Electroweak radiative corrections to neutral current Drell-Yan processes at
  hadron colliders}},  {\em Phys. Rev. D} {\bf 65} (2002) 033007,
  [\href{http://arxiv.org/abs/hep-ph/0108274}{{\tt hep-ph/0108274}}].

\bibitem{Zykunov:2005tc}
V.~Zykunov, {\it {Weak radiative corrections to Drell-Yan process for large
  invariant mass of di-lepton pair}},  {\em Phys. Rev. D} {\bf 75} (2007)
  073019, [\href{http://arxiv.org/abs/hep-ph/0509315}{{\tt hep-ph/0509315}}].

\bibitem{CarloniCalame:2007cd}
C.~Carloni~Calame, G.~Montagna, O.~Nicrosini, and A.~Vicini, {\it {Precision
  electroweak calculation of the production of a high transverse-momentum
  lepton pair at hadron colliders}},  {\em JHEP} {\bf 10} (2007) 109,
  [\href{http://arxiv.org/abs/0710.1722}{{\tt arXiv:0710.1722}}].

\bibitem{Arbuzov:2007db}
A.~Arbuzov, D.~Bardin, S.~Bondarenko, P.~Christova, L.~Kalinovskaya, G.~Nanava,
  and R.~Sadykov, {\it {One-loop corrections to the Drell--Yan process in SANC.
  (II). The Neutral current case}},  {\em Eur. Phys. J. C} {\bf 54} (2008)
  451--460, [\href{http://arxiv.org/abs/0711.0625}{{\tt arXiv:0711.0625}}].

\bibitem{Dittmaier:2009cr}
S.~Dittmaier and M.~Huber, {\it {Radiative corrections to the neutral-current
  Drell-Yan process in the Standard Model and its minimal supersymmetric
  extension}},  {\em JHEP} {\bf 01} (2010) 060,
  [\href{http://arxiv.org/abs/0911.2329}{{\tt arXiv:0911.2329}}].

\bibitem{Duhr:2020seh}
C.~Duhr, F.~Dulat, and B.~Mistlberger, {\it {Drell-Yan Cross Section to Third
  Order in the Strong Coupling Constant}},  {\em Phys. Rev. Lett.} {\bf 125}
  (2020), no.~17 172001, [\href{http://arxiv.org/abs/2001.07717}{{\tt
  arXiv:2001.07717}}].

\bibitem{Chen:2021vtu}
X.~Chen, T.~Gehrmann, N.~Glover, A.~Huss, T.-Z. Yang, and H.~X. Zhu, {\it
  {Di-lepton Rapidity Distribution in Drell-Yan Production to Third Order in
  QCD}},  \href{http://arxiv.org/abs/2107.09085}{{\tt arXiv:2107.09085}}.

\bibitem{Duhr:2020sdp}
C.~Duhr, F.~Dulat, and B.~Mistlberger, {\it {Charged current Drell-Yan
  production at N$^{3}$LO}},  {\em JHEP} {\bf 11} (2020) 143,
  [\href{http://arxiv.org/abs/2007.13313}{{\tt arXiv:2007.13313}}].

\bibitem{Camarda:2021ict}
S.~Camarda, L.~Cieri, and G.~Ferrera, {\it {Drell-Yan lepton-pair production:
  $q_T$ resummation at N$^3$LL accuracy and fiducial cross sections at
  N$^3$LO}},  \href{http://arxiv.org/abs/2103.04974}{{\tt arXiv:2103.04974}}.

\bibitem{Chen:2022cgv}
X.~Chen, T.~Gehrmann, E.~W.~N. Glover, A.~Huss, P.~F. Monni, E.~Re, L.~Rottoli,
  and P.~Torrielli, {\it {Third-Order Fiducial Predictions for Drell-Yan
  Production at the LHC}},  {\em Phys. Rev. Lett.} {\bf 128} (2022), no.~25
  252001, [\href{http://arxiv.org/abs/2203.01565}{{\tt arXiv:2203.01565}}].

\bibitem{Neumann:2022lft}
T.~Neumann and J.~Campbell, {\it {Fiducial Drell-Yan production at the LHC
  improved by transverse-momentum resummation at N4LLp+N3LO}},  {\em Phys. Rev.
  D} {\bf 107} (2023), no.~1 L011506,
  [\href{http://arxiv.org/abs/2207.07056}{{\tt arXiv:2207.07056}}].

\bibitem{Campbell:2023lcy}
J.~Campbell and T.~Neumann, {\it {Third order QCD predictions for fiducial
  W-boson production}},  {\em JHEP} {\bf 11} (2023) 127,
  [\href{http://arxiv.org/abs/2308.15382}{{\tt arXiv:2308.15382}}].

\bibitem{deFlorian:2018wcj}
D.~de~Florian, M.~Der, and I.~Fabre, {\it {QCD$\oplus$QED NNLO corrections to
  Drell Yan production}},  {\em Phys. Rev. D} {\bf 98} (2018), no.~9 094008,
  [\href{http://arxiv.org/abs/1805.12214}{{\tt arXiv:1805.12214}}].

\bibitem{Cieri:2020ikq}
L.~Cieri, D.~de~Florian, M.~Der, and J.~Mazzitelli, {\it {Mixed
  QCD\ensuremath{\otimes}QED corrections to exclusive Drell Yan production
  using the q$_{T}$ -subtraction method}},  {\em JHEP} {\bf 09} (2020) 155,
  [\href{http://arxiv.org/abs/2005.01315}{{\tt arXiv:2005.01315}}].

\bibitem{Delto:2019ewv}
M.~Delto, M.~Jaquier, K.~Melnikov, and R.~R\"ontsch, {\it {Mixed
  QCD$\otimes$QED corrections to on-shell $Z$ boson production at the LHC}},
  {\em JHEP} {\bf 01} (2020) 043, [\href{http://arxiv.org/abs/1909.08428}{{\tt
  arXiv:1909.08428}}].

\bibitem{Bonciani:2016wya}
R.~Bonciani, F.~Buccioni, R.~Mondini, and A.~Vicini, {\it {Double-real
  corrections at $\mathcal{O}(\alpha \alpha_s)$ to single gauge boson
  production}},  {\em Eur. Phys. J. C} {\bf 77} (2017), no.~3 187,
  [\href{http://arxiv.org/abs/1611.00645}{{\tt arXiv:1611.00645}}].

\bibitem{Bonciani:2019nuy}
R.~Bonciani, F.~Buccioni, N.~Rana, I.~Triscari, and A.~Vicini, {\it {NNLO
  QCD$\times$EW corrections to Z production in the $q\bar{q}$ channel}},  {\em
  Phys. Rev. D} {\bf 101} (2020), no.~3 031301,
  [\href{http://arxiv.org/abs/1911.06200}{{\tt arXiv:1911.06200}}].

\bibitem{Bonciani:2020tvf}
R.~Bonciani, F.~Buccioni, N.~Rana, and A.~Vicini, {\it {NNLO QCD$\times$EW
  corrections to on-shell $Z$ production}},  {\em Phys. Rev. Lett.} {\bf 125}
  (2020), no.~23 232004, [\href{http://arxiv.org/abs/2007.06518}{{\tt
  arXiv:2007.06518}}].

\bibitem{Buccioni:2020cfi}
F.~Buccioni, F.~Caola, M.~Delto, M.~Jaquier, K.~Melnikov, and R.~R\"ontsch,
  {\it {Mixed QCD-electroweak corrections to on-shell Z production at the
  LHC}},  {\em Phys. Lett. B} {\bf 811} (2020) 135969,
  [\href{http://arxiv.org/abs/2005.10221}{{\tt arXiv:2005.10221}}].

\bibitem{Behring:2020cqi}
A.~Behring, F.~Buccioni, F.~Caola, M.~Delto, M.~Jaquier, K.~Melnikov, and
  R.~R\"ontsch, {\it {Mixed QCD-electroweak corrections to $W$-boson production
  in hadron collisions}},  {\em Phys. Rev. D} {\bf 103} (2021), no.~1 013008,
  [\href{http://arxiv.org/abs/2009.10386}{{\tt arXiv:2009.10386}}].

\bibitem{Bonciani:2021iis}
R.~Bonciani, F.~Buccioni, N.~Rana, and A.~Vicini, {\it {On-shell Z boson
  production at hadron colliders through
  \ensuremath{\mathscr{O}}(\ensuremath{\alpha}\ensuremath{\alpha}$_{s}$)}},
  {\em JHEP} {\bf 02} (2022) 095, [\href{http://arxiv.org/abs/2111.12694}{{\tt
  arXiv:2111.12694}}].

\bibitem{Denner:2019vbn}
A.~Denner and S.~Dittmaier, {\it {Electroweak Radiative Corrections for
  Collider Physics}},  {\em Phys. Rept.} {\bf 864} (2020) 1--163,
  [\href{http://arxiv.org/abs/1912.06823}{{\tt arXiv:1912.06823}}].

\bibitem{Dittmaier:2014qza}
S.~Dittmaier, A.~Huss, and C.~Schwinn, {\it {Mixed QCD-electroweak
  $\mathcal{O}(\alpha_s\alpha)$ corrections to Drell-Yan processes in the
  resonance region: pole approximation and non-factorizable corrections}},
  {\em Nucl. Phys. B} {\bf 885} (2014) 318--372,
  [\href{http://arxiv.org/abs/1403.3216}{{\tt arXiv:1403.3216}}].

\bibitem{Dittmaier:2015rxo}
S.~Dittmaier, A.~Huss, and C.~Schwinn, {\it {Dominant mixed QCD-electroweak
  O($\alpha$$_s$$\alpha$) corrections to Drell\textendash{}Yan processes in the
  resonance region}},  {\em Nucl. Phys. B} {\bf 904} (2016) 216--252,
  [\href{http://arxiv.org/abs/1511.08016}{{\tt arXiv:1511.08016}}].

\bibitem{Dittmaier:2024row}
S.~Dittmaier, A.~Huss, and J.~Schwarz, {\it {Mixed NNLO QCD \texttimes{}
  electroweak corrections to single-Z production in pole approximation:
  differential distributions and forward-backward asymmetry}},  {\em JHEP} {\bf
  05} (2024) 170, [\href{http://arxiv.org/abs/2401.15682}{{\tt
  arXiv:2401.15682}}].

\bibitem{Dittmaier:2020vra}
S.~Dittmaier, T.~Schmidt, and J.~Schwarz, {\it {Mixed NNLO
  QCD$\times$electroweak corrections of $\mathcal{O}(N_f \alpha_s \alpha)$ to
  single-W/Z production at the LHC}},  {\em JHEP} {\bf 12} (2020) 201,
  [\href{http://arxiv.org/abs/2009.02229}{{\tt arXiv:2009.02229}}].

\bibitem{Buonocore:2021rxx}
L.~Buonocore, M.~Grazzini, S.~Kallweit, C.~Savoini, and F.~Tramontano, {\it
  {Mixed QCD-EW corrections to $pp\!\to\!\ell\nu_\ell\!+\!X$ at the LHC}},
  {\em Phys. Rev. D} {\bf 103} (2021) 114012,
  [\href{http://arxiv.org/abs/2102.12539}{{\tt arXiv:2102.12539}}].

\bibitem{Armadillo:2024nwk}
T.~Armadillo, R.~Bonciani, S.~Devoto, N.~Rana, and A.~Vicini, {\it {Two-loop
  mixed QCD-EW corrections to charged current Drell-Yan}},  {\em JHEP} {\bf 07}
  (2024) 265, [\href{http://arxiv.org/abs/2405.00612}{{\tt arXiv:2405.00612}}].

\bibitem{Bonciani:2021zzf}
R.~Bonciani, L.~Buonocore, M.~Grazzini, S.~Kallweit, N.~Rana, F.~Tramontano,
  and A.~Vicini, {\it {Mixed Strong-Electroweak Corrections to the Drell-Yan
  Process}},  {\em Phys. Rev. Lett.} {\bf 128} (2022), no.~1 012002,
  [\href{http://arxiv.org/abs/2106.11953}{{\tt arXiv:2106.11953}}].

\bibitem{Buccioni:2022kgy}
F.~Buccioni, F.~Caola, H.~A. Chawdhry, F.~Devoto, M.~Heller, A.~von Manteuffel,
  K.~Melnikov, R.~R\"ontsch, and C.~Signorile-Signorile, {\it {Mixed
  QCD-electroweak corrections to dilepton production at the LHC in the high
  invariant mass region}},  \href{http://arxiv.org/abs/2203.11237}{{\tt
  arXiv:2203.11237}}.

\bibitem{Armadillo:2022bgm}
T.~Armadillo, R.~Bonciani, S.~Devoto, N.~Rana, and A.~Vicini, {\it {Two-loop
  mixed QCD-EW corrections to neutral current Drell-Yan}},  {\em JHEP} {\bf 05}
  (2022) 072, [\href{http://arxiv.org/abs/2201.01754}{{\tt arXiv:2201.01754}}].

\bibitem{Bonciani:2016ypc}
R.~Bonciani, S.~Di~Vita, P.~Mastrolia, and U.~Schubert, {\it {Two-Loop Master
  Integrals for the mixed EW-QCD virtual corrections to Drell-Yan scattering}},
   {\em JHEP} {\bf 09} (2016) 091, [\href{http://arxiv.org/abs/1604.08581}{{\tt
  arXiv:1604.08581}}].

\bibitem{Moriello:2019yhu}
F.~Moriello, {\it {Generalised power series expansions for the elliptic planar
  families of Higgs + jet production at two loops}},  {\em JHEP} {\bf 01}
  (2020) 150, [\href{http://arxiv.org/abs/1907.13234}{{\tt arXiv:1907.13234}}].

\bibitem{Hidding:2020ytt}
M.~Hidding, {\it {DiffExp, a Mathematica package for computing Feynman
  integrals in terms of one-dimensional series expansions}},  {\em Comput.
  Phys. Commun.} {\bf 269} (2021) 108125,
  [\href{http://arxiv.org/abs/2006.05510}{{\tt arXiv:2006.05510}}].

\bibitem{Armadillo:2022ugh}
T.~Armadillo, R.~Bonciani, S.~Devoto, N.~Rana, and A.~Vicini, {\it {Evaluation
  of Feynman integrals with arbitrary complex masses via series expansions}},
  {\em Comput. Phys. Commun.} {\bf 282} (2023) 108545,
  [\href{http://arxiv.org/abs/2205.03345}{{\tt arXiv:2205.03345}}].

\bibitem{Heller:2020owb}
M.~Heller, A.~von Manteuffel, R.~M. Schabinger, and H.~Spiesberger, {\it {Mixed
  EW-QCD two-loop amplitudes for $q\bar{q} \to \ell^+\ell^-$ and $\gamma_5$
  scheme independence of multi-loop corrections}},  {\em JHEP} {\bf 05} (2021)
  213, [\href{http://arxiv.org/abs/2012.05918}{{\tt arXiv:2012.05918}}].

\bibitem{Heller:2019gkq}
M.~Heller, A.~von Manteuffel, and R.~M. Schabinger, {\it {Multiple
  polylogarithms with algebraic arguments and the two-loop EW-QCD Drell-Yan
  master integrals}},  {\em Phys. Rev. D} {\bf 102} (2020), no.~1 016025,
  [\href{http://arxiv.org/abs/1907.00491}{{\tt arXiv:1907.00491}}].

\bibitem{Hasan:2020vwn}
S.~M. Hasan and U.~Schubert, {\it {Master Integrals for the mixed QCD-QED
  corrections to the Drell-Yan production of a massive lepton pair}},  {\em
  JHEP} {\bf 11} (2020) 107, [\href{http://arxiv.org/abs/2004.14908}{{\tt
  arXiv:2004.14908}}].

\bibitem{Balossini:2008cs}
G.~Balossini, G.~Montagna, C.~M. Carloni~Calame, M.~Moretti, M.~Treccani,
  O.~Nicrosini, F.~Piccinini, and A.~Vicini, {\it {Electroweak \& QCD
  corrections to Drell Yan processes}},  {\em Acta Phys. Polon. B} {\bf 39}
  (2008) 1675, [\href{http://arxiv.org/abs/0805.1129}{{\tt arXiv:0805.1129}}].

\bibitem{Balossini:2009sa}
G.~Balossini, G.~Montagna, C.~M. Carloni~Calame, M.~Moretti, O.~Nicrosini,
  F.~Piccinini, M.~Treccani, and A.~Vicini, {\it {Combination of electroweak
  and QCD corrections to single W production at the Fermilab Tevatron and the
  CERN LHC}},  {\em JHEP} {\bf 01} (2010) 013,
  [\href{http://arxiv.org/abs/0907.0276}{{\tt arXiv:0907.0276}}].

\bibitem{Grazzini:2019jkl}
M.~Grazzini, S.~Kallweit, J.~M. Lindert, S.~Pozzorini, and M.~Wiesemann, {\it
  {NNLO QCD + NLO EW with Matrix+OpenLoops: precise predictions for
  vector-boson pair production}},  {\em JHEP} {\bf 02} (2020) 087,
  [\href{http://arxiv.org/abs/1912.00068}{{\tt arXiv:1912.00068}}].

\bibitem{deFlorian:2015ujt}
D.~de~Florian, G.~F.~R. Sborlini, and G.~Rodrigo, {\it {QED corrections to the
  Altarelli\textendash{}Parisi splitting functions}},  {\em Eur. Phys. J. C}
  {\bf 76} (2016), no.~5 282, [\href{http://arxiv.org/abs/1512.00612}{{\tt
  arXiv:1512.00612}}].

\bibitem{deFlorian:2016gvk}
D.~de~Florian, G.~F.~R. Sborlini, and G.~Rodrigo, {\it {Two-loop QED
  corrections to the Altarelli-Parisi splitting functions}},  {\em JHEP} {\bf
  10} (2016) 056, [\href{http://arxiv.org/abs/1606.02887}{{\tt
  arXiv:1606.02887}}].

\bibitem{Xie:2021equ}
{\bf CTEQ-TEA} Collaboration, K.~Xie, T.~J. Hobbs, T.-J. Hou, C.~Schmidt,
  M.~Yan, and C.~P. Yuan, {\it {Photon PDF within the CT18 global analysis}},
  {\em Phys. Rev. D} {\bf 105} (2022), no.~5 054006,
  [\href{http://arxiv.org/abs/2106.10299}{{\tt arXiv:2106.10299}}].

\bibitem{Cridge:2023ryv}
T.~Cridge, L.~A. Harland-Lang, and R.~S. Thorne, {\it {Combining QED and
  Approximate N${}^3$LO QCD Corrections in a Global PDF Fit: MSHT20qed\_an3lo
  PDFs}},  \href{http://arxiv.org/abs/2312.07665}{{\tt arXiv:2312.07665}}.

\bibitem{Barontini:2024dyb}
A.~Barontini, N.~Laurenti, and J.~Rojo, {\it {NNPDF4.0 aN$^3$LO PDFs with QED
  corrections}},  in {\em {31st International Workshop on Deep-Inelastic
  Scattering and Related Subjects}}, 6, 2024.
\newblock \href{http://arxiv.org/abs/2406.01779}{{\tt arXiv:2406.01779}}.

\bibitem{Mangano:2016jyj}
M.~L. Mangano et~al., {\it {Physics at a 100 TeV pp Collider: Standard Model
  Processes}},  \href{http://arxiv.org/abs/1607.01831}{{\tt arXiv:1607.01831}}.

\bibitem{Manohar:2016nzj}
A.~Manohar, P.~Nason, G.~P. Salam, and G.~Zanderighi, {\it {How bright is the
  proton? A precise determination of the photon parton distribution function}},
   {\em Phys. Rev. Lett.} {\bf 117} (2016), no.~24 242002,
  [\href{http://arxiv.org/abs/1607.04266}{{\tt arXiv:1607.04266}}].

\bibitem{Manohar:2017eqh}
A.~V. Manohar, P.~Nason, G.~P. Salam, and G.~Zanderighi, {\it {The Photon
  Content of the Proton}},  {\em JHEP} {\bf 12} (2017) 046,
  [\href{http://arxiv.org/abs/1708.01256}{{\tt arXiv:1708.01256}}].

\bibitem{Ball:2022qtp}
R.~D. Ball, A.~Candido, S.~Forte, F.~Hekhorn, E.~R. Nocera, J.~Rojo, and
  C.~Schwan, {\it {Parton distributions and new physics searches: the
  Drell\textendash{}Yan forward\textendash{}backward asymmetry as a case
  study}},  {\em Eur. Phys. J. C} {\bf 82} (2022), no.~12 1160,
  [\href{http://arxiv.org/abs/2209.08115}{{\tt arXiv:2209.08115}}].

\bibitem{Barze:2013fru}
L.~Barze, G.~Montagna, P.~Nason, O.~Nicrosini, F.~Piccinini, and A.~Vicini,
  {\it {Neutral current Drell-Yan with combined QCD and electroweak corrections
  in the POWHEG BOX}},  {\em Eur. Phys. J. C} {\bf 73} (2013), no.~6 2474,
  [\href{http://arxiv.org/abs/1302.4606}{{\tt arXiv:1302.4606}}].

\bibitem{Kallweit:2015dum}
S.~Kallweit, J.~M. Lindert, P.~Maierh\"ofer, S.~Pozzorini, and M.~Sch\"onherr,
  {\it {NLO QCD+EW predictions for V + jets including off-shell vector-boson
  decays and multijet merging}},  {\em JHEP} {\bf 04} (2016) 021,
  [\href{http://arxiv.org/abs/1511.08692}{{\tt arXiv:1511.08692}}].

\bibitem{Andersen:2016qtm}
J.~R. Andersen et~al., {\it {Les Houches 2015: Physics at TeV Colliders
  Standard Model Working Group Report}},  in {\em {9th Les Houches Workshop on
  Physics at TeV Colliders}}, 5, 2016.
\newblock \href{http://arxiv.org/abs/1605.04692}{{\tt arXiv:1605.04692}}.

\bibitem{Alioli:2016fum}
S.~Alioli et~al., {\it {Precision studies of observables in $p p \rightarrow W
  \rightarrow l\nu _l$ and $ pp \rightarrow \gamma ,Z \rightarrow l^+ l^-$
  processes at the LHC}},  {\em Eur. Phys. J. C} {\bf 77} (2017), no.~5 280,
  [\href{http://arxiv.org/abs/1606.02330}{{\tt arXiv:1606.02330}}].

\bibitem{Cascioli:2011va}
F.~Cascioli, P.~Maierh{\"o}fer, and S.~Pozzorini, {\it {Scattering Amplitudes
  with Open Loops}},  {\em Phys. Rev. Lett.} {\bf 108} (2012) 111601,
  [\href{http://arxiv.org/abs/1111.5206}{{\tt arXiv:1111.5206}}].

\bibitem{Buccioni:2017yxi}
F.~Buccioni, S.~Pozzorini, and M.~Zoller, {\it {On-the-fly reduction of open
  loops}},  {\em Eur. Phys. J. C} {\bf 78} (2018), no.~1 70,
  [\href{http://arxiv.org/abs/1710.11452}{{\tt arXiv:1710.11452}}].

\bibitem{Buccioni:2019sur}
F.~Buccioni, J.-N. Lang, J.~M. Lindert, P.~Maierh{\"o}fer, S.~Pozzorini,
  H.~Zhang, and M.~F. Zoller, {\it {OpenLoops 2}},  {\em Eur. Phys. J. C} {\bf
  79} (2019), no.~10 866, [\href{http://arxiv.org/abs/1907.13071}{{\tt
  arXiv:1907.13071}}].

\bibitem{Actis:2016mpe}
S.~Actis, A.~Denner, L.~Hofer, J.-N. Lang, A.~Scharf, and S.~Uccirati, {\it
  {RECOLA: REcursive Computation of One-Loop Amplitudes}},  {\em Comput. Phys.
  Commun.} {\bf 214} (2017) 140--173,
  [\href{http://arxiv.org/abs/1605.01090}{{\tt arXiv:1605.01090}}].

\bibitem{Denner:2017wsf}
A.~Denner, J.-N. Lang, and S.~Uccirati, {\it {Recola2: REcursive Computation of
  One-Loop Amplitudes 2}},  {\em Comput. Phys. Commun.} {\bf 224} (2018)
  346--361, [\href{http://arxiv.org/abs/1711.07388}{{\tt arXiv:1711.07388}}].

\bibitem{Denner:2016kdg}
A.~Denner, S.~Dittmaier, and L.~Hofer, {\it {Collier: a fortran-based Complex
  One-Loop LIbrary in Extended Regularizations}},  {\em Comput. Phys. Commun.}
  {\bf 212} (2017) 220--238, [\href{http://arxiv.org/abs/1604.06792}{{\tt
  arXiv:1604.06792}}].

\bibitem{Maierhoefer:2017hyi}
P.~Maierh\"ofer, J.~Usovitsch, and P.~Uwer, {\it {Kira\textemdash{}A Feynman
  integral reduction program}},  {\em Comput. Phys. Commun.} {\bf 230} (2018)
  99--112, [\href{http://arxiv.org/abs/1705.05610}{{\tt arXiv:1705.05610}}].

\bibitem{Denner:2005fg}
A.~Denner, S.~Dittmaier, M.~Roth, and L.~H. Wieders, {\it {Electroweak
  corrections to charged-current e+ e- ---\ensuremath{>} 4 fermion processes:
  Technical details and further results}},  {\em Nucl. Phys. B} {\bf 724}
  (2005) 247--294, [\href{http://arxiv.org/abs/hep-ph/0505042}{{\tt
  hep-ph/0505042}}]. [Erratum: Nucl.Phys.B 854, 504--507 (2012)].

\bibitem{Catani:2007vq}
S.~Catani and M.~Grazzini, {\it {An NNLO subtraction formalism in hadron
  collisions and its application to Higgs boson production at the LHC}},  {\em
  Phys. Rev. Lett.} {\bf 98} (2007) 222002,
  [\href{http://arxiv.org/abs/hep-ph/0703012}{{\tt hep-ph/0703012}}].

\bibitem{Catani:2019iny}
S.~Catani, S.~Devoto, M.~Grazzini, S.~Kallweit, J.~Mazzitelli, and H.~Sargsyan,
  {\it {Top-quark pair hadroproduction at next-to-next-to-leading order in
  QCD}},  {\em Phys. Rev. D} {\bf 99} (2019), no.~5 051501,
  [\href{http://arxiv.org/abs/1901.04005}{{\tt arXiv:1901.04005}}].

\bibitem{Catani:2019hip}
S.~Catani, S.~Devoto, M.~Grazzini, S.~Kallweit, and J.~Mazzitelli, {\it
  {Top-quark pair production at the LHC: Fully differential QCD predictions at
  NNLO}},  {\em JHEP} {\bf 07} (2019) 100,
  [\href{http://arxiv.org/abs/1906.06535}{{\tt arXiv:1906.06535}}].

\bibitem{Catani:2020kkl}
S.~Catani, S.~Devoto, M.~Grazzini, S.~Kallweit, and J.~Mazzitelli, {\it
  {Bottom-quark production at hadron colliders: fully differential predictions
  in NNLO QCD}},  {\em JHEP} {\bf 03} (2021) 029,
  [\href{http://arxiv.org/abs/2010.11906}{{\tt arXiv:2010.11906}}].

\bibitem{Buonocore:2019puv}
L.~Buonocore, M.~Grazzini, and F.~Tramontano, {\it {The $q_T$ subtraction
  method: electroweak corrections and power suppressed contributions}},  {\em
  Eur. Phys. J. C} {\bf 80} (2020), no.~3 254,
  [\href{http://arxiv.org/abs/1911.10166}{{\tt arXiv:1911.10166}}].

\bibitem{Catani:1996jh}
S.~Catani and M.~H. Seymour, {\it {The Dipole formalism for the calculation of
  QCD jet cross-sections at next-to-leading order}},  {\em Phys. Lett. B} {\bf
  378} (1996) 287--301, [\href{http://arxiv.org/abs/hep-ph/9602277}{{\tt
  hep-ph/9602277}}].

\bibitem{Catani:1996vz}
S.~Catani and M.~H. Seymour, {\it {A General algorithm for calculating jet
  cross-sections in NLO QCD}},  {\em Nucl. Phys. B} {\bf 485} (1997) 291--419,
  [\href{http://arxiv.org/abs/hep-ph/9605323}{{\tt hep-ph/9605323}}]. [Erratum:
  Nucl.Phys.B 510, 503--504 (1998)].

\bibitem{Catani:2002hc}
S.~Catani, S.~Dittmaier, M.~H. Seymour, and Z.~Trocsanyi, {\it {The Dipole
  formalism for next-to-leading order QCD calculations with massive partons}},
  {\em Nucl. Phys. B} {\bf 627} (2002) 189--265,
  [\href{http://arxiv.org/abs/hep-ph/0201036}{{\tt hep-ph/0201036}}].

\bibitem{Kallweit:2017khh}
S.~Kallweit, J.~M. Lindert, S.~Pozzorini, and M.~Sch\"onherr, {\it {NLO QCD+EW
  predictions for $2\ell2\nu$ diboson signatures at the LHC}},  {\em JHEP} {\bf
  11} (2017) 120, [\href{http://arxiv.org/abs/1705.00598}{{\tt
  arXiv:1705.00598}}].

\bibitem{Dittmaier:1999mb}
S.~Dittmaier, {\it {A General approach to photon radiation off fermions}},
  {\em Nucl. Phys. B} {\bf 565} (2000) 69--122,
  [\href{http://arxiv.org/abs/hep-ph/9904440}{{\tt hep-ph/9904440}}].

\bibitem{Dittmaier:2008md}
S.~Dittmaier, A.~Kabelschacht, and T.~Kasprzik, {\it {Polarized QED splittings
  of massive fermions and dipole subtraction for non-collinear-safe
  observables}},  {\em Nucl. Phys. B} {\bf 800} (2008) 146--189,
  [\href{http://arxiv.org/abs/0802.1405}{{\tt arXiv:0802.1405}}].

\bibitem{Gehrmann:2010ry}
T.~Gehrmann and N.~Greiner, {\it {Photon Radiation with MadDipole}},  {\em
  JHEP} {\bf 12} (2010) 050, [\href{http://arxiv.org/abs/1011.0321}{{\tt
  arXiv:1011.0321}}].

\bibitem{Schonherr:2017qcj}
M.~Sch\"onherr, {\it {An automated subtraction of NLO EW infrared
  divergences}},  {\em Eur. Phys. J. C} {\bf 78} (2018), no.~2 119,
  [\href{http://arxiv.org/abs/1712.07975}{{\tt arXiv:1712.07975}}].

\bibitem{Grazzini:2017mhc}
M.~Grazzini, S.~Kallweit, and M.~Wiesemann, {\it {Fully differential NNLO
  computations with MATRIX}},  {\em Eur. Phys. J. C} {\bf 78} (2018), no.~7
  537, [\href{http://arxiv.org/abs/1711.06631}{{\tt arXiv:1711.06631}}].

\bibitem{Catani:2022mfv}
S.~Catani, S.~Devoto, M.~Grazzini, S.~Kallweit, J.~Mazzitelli, and C.~Savoini,
  {\it {Higgs Boson Production in Association with a Top-Antitop Quark Pair in
  Next-to-Next-to-Leading Order QCD}},  {\em Phys. Rev. Lett.} {\bf 130}
  (2023), no.~11 111902, [\href{http://arxiv.org/abs/2210.07846}{{\tt
  arXiv:2210.07846}}].

\bibitem{Buonocore:2022pqq}
L.~Buonocore, S.~Devoto, S.~Kallweit, J.~Mazzitelli, L.~Rottoli, and
  C.~Savoini, {\it {Associated production of a W boson and massive bottom
  quarks at next-to-next-to-leading order in QCD}},  {\em Phys. Rev. D} {\bf
  107} (2023), no.~7 074032, [\href{http://arxiv.org/abs/2212.04954}{{\tt
  arXiv:2212.04954}}].

\bibitem{Buonocore:2023ljm}
L.~Buonocore, S.~Devoto, M.~Grazzini, S.~Kallweit, J.~Mazzitelli, L.~Rottoli,
  and C.~Savoini, {\it {Precise Predictions for the Associated Production of a
  W Boson with a Top-Antitop Quark Pair at the LHC}},  {\em Phys. Rev. Lett.}
  {\bf 131} (2023), no.~23 231901, [\href{http://arxiv.org/abs/2306.16311}{{\tt
  arXiv:2306.16311}}].

\bibitem{Devoto:2024nhl}
S.~Devoto, M.~Grazzini, S.~Kallweit, J.~Mazzitelli, and C.~Savoini, {\it
  {Precise predictions for $t \bar t H$ production at the LHC: inclusive cross
  section and differential distributions}},
  \href{http://arxiv.org/abs/2411.15340}{{\tt arXiv:2411.15340}}.

\bibitem{Catani:2023tby}
S.~Catani, S.~Devoto, M.~Grazzini, and J.~Mazzitelli, {\it {Soft-parton
  contributions to heavy-quark production at low transverse momentum}},  {\em
  JHEP} {\bf 04} (2023) 144, [\href{http://arxiv.org/abs/2301.11786}{{\tt
  arXiv:2301.11786}}].

\bibitem{Bertone:2017bme}
{\bf NNPDF} Collaboration, V.~Bertone, S.~Carrazza, N.~P. Hartland, and
  J.~Rojo, {\it {Illuminating the photon content of the proton within a global
  PDF analysis}},  {\em SciPost Phys.} {\bf 5} (2018), no.~1 008,
  [\href{http://arxiv.org/abs/1712.07053}{{\tt arXiv:1712.07053}}].

\bibitem{Salam:2021tbm}
G.~P. Salam and E.~Slade, {\it {Cuts for two-body decays at colliders}},  {\em
  JHEP} {\bf 11} (2021) 220, [\href{http://arxiv.org/abs/2106.08329}{{\tt
  arXiv:2106.08329}}].

\bibitem{NNPDF:2024dpb}
{\bf NNPDF} Collaboration, R.~D. Ball et~al., {\it {Determination of the theory
  uncertainties from missing higher orders on NNLO parton distributions with
  percent accuracy}},  {\em Eur. Phys. J. C} {\bf 84} (2024), no.~5 517,
  [\href{http://arxiv.org/abs/2401.10319}{{\tt arXiv:2401.10319}}].

\bibitem{Collins:1977iv}
J.~C. Collins and D.~E. Soper, {\it {Angular Distribution of Dileptons in
  High-Energy Hadron Collisions}},  {\em Phys. Rev. D} {\bf 16} (1977) 2219.

\bibitem{CMS:2021ctt}
{\bf CMS} Collaboration, A.~M. Sirunyan et~al., {\it {Search for resonant and
  nonresonant new phenomena in high-mass dilepton final states at $ \sqrt{s} $
  = 13 TeV}},  {\em JHEP} {\bf 07} (2021) 208,
  [\href{http://arxiv.org/abs/2103.02708}{{\tt arXiv:2103.02708}}].

\bibitem{Hammou:2023heg}
E.~Hammou, Z.~Kassabov, M.~Madigan, M.~L. Mangano, L.~Mantani, J.~Moore, M.~M.
  Alvarado, and M.~Ubiali, {\it {Hide and seek: how PDFs can conceal new
  physics}},  {\em JHEP} {\bf 11} (2023) 090,
  [\href{http://arxiv.org/abs/2307.10370}{{\tt arXiv:2307.10370}}].

\bibitem{Amoroso:2023uux}
S.~Amoroso, M.~Chiesa, C.~L. Del~Pio, K.~Lipka, F.~Piccinini, F.~Vazzoler, and
  A.~Vicini, {\it {Probing the weak mixing angle at high energies at the LHC
  and HL-LHC}},  {\em Phys. Lett. B} {\bf 844} (2023) 138103,
  [\href{http://arxiv.org/abs/2302.10782}{{\tt arXiv:2302.10782}}].

\bibitem{Baur:1997wa}
U.~Baur, S.~Keller, and W.~K. Sakumoto, {\it {QED radiative corrections to $Z$
  boson production and the forward backward asymmetry at hadron colliders}},
  {\em Phys. Rev. D} {\bf 57} (1998) 199--215,
  [\href{http://arxiv.org/abs/hep-ph/9707301}{{\tt hep-ph/9707301}}].

\bibitem{ALEPH:2005ab}
{\bf ALEPH, DELPHI, L3, OPAL, SLD, LEP Electroweak Working Group, SLD
  Electroweak Group, SLD Heavy Flavour Group} Collaboration, S.~Schael et~al.,
  {\it {Precision electroweak measurements on the $Z$ resonance}},  {\em Phys.
  Rept.} {\bf 427} (2006) 257--454,
  [\href{http://arxiv.org/abs/hep-ex/0509008}{{\tt hep-ex/0509008}}].

\bibitem{CMS:2018ktx}
{\bf CMS} Collaboration, A.~M. Sirunyan et~al., {\it {Measurement of the weak
  mixing angle using the forward-backward asymmetry of Drell-Yan events in pp
  collisions at 8 TeV}},  {\em Eur. Phys. J. C} {\bf 78} (2018), no.~9 701,
  [\href{http://arxiv.org/abs/1806.00863}{{\tt arXiv:1806.00863}}].

\bibitem{ATLAS:2015ihy}
{\bf ATLAS} Collaboration, G.~Aad et~al., {\it {Measurement of the
  forward-backward asymmetry of electron and muon pair-production in $pp$
  collisions at $\sqrt{s}$ = 7 TeV with the ATLAS detector}},  {\em JHEP} {\bf
  09} (2015) 049, [\href{http://arxiv.org/abs/1503.03709}{{\tt
  arXiv:1503.03709}}].

\bibitem{Chiesa:2019nqb}
M.~Chiesa, F.~Piccinini, and A.~Vicini, {\it {Direct determination of $\sin^2
  \theta^\ell_{eff}$ at hadron colliders}},  {\em Phys. Rev. D} {\bf 100}
  (2019), no.~7 071302, [\href{http://arxiv.org/abs/1906.11569}{{\tt
  arXiv:1906.11569}}].

\bibitem{ParticleDataGroup:2024cfk}
{\bf Particle Data Group} Collaboration, S.~Navas et~al., {\it {Review of
  particle physics}},  {\em Phys. Rev. D} {\bf 110} (2024), no.~3 030001.

\bibitem{Cacciari:1998it}
M.~Cacciari, M.~Greco, and P.~Nason, {\it {The P(T) spectrum in heavy flavor
  hadroproduction}},  {\em JHEP} {\bf 05} (1998) 007,
  [\href{http://arxiv.org/abs/hep-ph/9803400}{{\tt hep-ph/9803400}}].

\bibitem{Collins:1986mp}
J.~C. Collins and W.-K. Tung, {\it {Calculating Heavy Quark Distributions}},
  {\em Nucl. Phys. B} {\bf 278} (1986) 934.

\end{thebibliography}\endgroup
\end{document}